\documentclass[fleqn,usenatbib]{mnras}

\usepackage{newtxtext,newtxmath}

\usepackage[T1]{fontenc}

\DeclareRobustCommand{\VAN}[3]{#2}
\let\VANthebibliography\thebibliography
\def\thebibliography{\DeclareRobustCommand{\VAN}[3]{##3}\VANthebibliography}

\usepackage{graphicx}	
\usepackage{amsmath}	
\usepackage{caption}
\usepackage{subcaption}
\usepackage{orcidlink}

\title[]{Modeling Strong Lenses from Wide-Field Ground-Based Observations in KiDS and GAMA}

\author[S. Knabel]{Shawn Knabel$^{1,2}$\orcidlink{0000-0001-5110-6241}, 
B. W. Holwerda$^{1}$\orcidlink{0000-0002-4884-6756}, J. Nightingale$^{3}$\orcidlink{0000-0002-8987-7401}, 
T. Treu$^{2}$\orcidlink{0000-0002-8460-0390},
\newauthor 
M. Bilicki$^{4}$\orcidlink{0000-0002-3910-5809}, 
S. Brough$^{5}$\orcidlink{0000-0002-9796-1363}, 
S. Driver$^{6}$\orcidlink{0000-0001-9491-7327}, 
L. Finnerty$^{2}$\orcidlink{0000-0002-1392-0768}
L. Haberzettl$^{1}$\orcidlink{0000-0001-5620-3024}, 
\newauthor
S. Hegde$^{2}$\orcidlink{0000-0002-9370-8061},
A. M. Hopkins$^{7}$\orcidlink{0000-0002-6097-2747}, 
K. Kuijken$^{8}$\orcidlink{0000-0002-3827-0175}, 
J. Liske$^{9}$\orcidlink{0000-0001-7542-2927}, 
K. A. Pimbblet$^{10}$\orcidlink{0000-0002-3963-3919}, 
\newauthor
R. C. Steele$^{1}$\orcidlink{0000-0001-9537-5814}, and
A. H. Wright$^{11}$\orcidlink{0000-0001-7363-7932} \\
$^{1}$ University of Louisville, Department of Physics and Astronomy, 102 Natural Science Building, 40292 KY Louisville, USA.\\
$^{2}$ University of California-Los Angeles, Department of Physics and Astronomy, PAB, 430 Portola
Plaza, Box 951547, Los Angeles, CA 90095-1547, USA \\
$^{3}$ Durham University, Institute for Computational Cosmology, Stockton Rd, Durham DH1 3LE, United Kingdom \\
$^{4}$ Center for Theoretical Physics, Polish Academy of Sciences, al. Lotników 32/46, 02-668 Warsaw, Poland \\
$^{5}$ School of Physics, University of New South Wales, NSW 2052, Australia \\ 
$^{6}$ International Centre for Radio Astronomy Research (ICRAR), University of Western Australia, Crawley, Australia, WA 6009 \\
$^{7}$ Australian Astronomical Optics, Macquarie University, 105 Delhi Rd, North Ryde, NSW 2113, Australia \\
$^{8}$ Leiden Observatory, Leiden University, P.O. Box 9513, 2300RA Leiden, the Netherlands \\
$^{9}$ Universit\"at Hamburg, Hamburg Sternwarte, Gojenbergsweg 112, 21029 Hamburg, Germany \\
$^{10}$ E.A.Milne Centre for Astrophysics, University of Hull, Cottingham Road, Kingston-upon-Hull, HU6 7RX, UK \\
$^{11}$ German Center for Cosmological Lensing (GCCL), Astronomisches Institut, Ruhr-Universität Bochum, Universitätsstraße 150, 44780, Bochum, Germany \\
}

\date{This is a pre-copyedited, author-produced PDF of an article accepted for publication in Monthly Notices of the Royal Astronomical Society following peer review. }

\pubyear{2021}

\begin{document}
\label{firstpage}
\pagerange{\pageref{firstpage}--\pageref{lastpage}}
\maketitle

\begin{abstract}

Despite the success of galaxy-scale strong gravitational lens studies with Hubble-quality imaging, the number of well-studied strong lenses remains small. As a result, robust comparisons of the lens models to theoretical predictions are difficult. This motivates our application of automated Bayesian lens modeling methods to observations from public data releases of overlapping large ground-based imaging and spectroscopic surveys: Kilo-Degree Survey (KiDS) and Galaxy and Mass Assembly (GAMA), respectively. We use the open-source lens modeling software {\sc PyAutoLens} to perform our analysis. We demonstrate the feasibility of strong lens modeling with large-survey data at lower resolution as a complementary avenue to studies that utilize more time-consuming and expensive observations of individual lenses at higher resolution. We discuss advantages and challenges, with special consideration given to determining background source redshifts from single-aperture spectra and to disentangling foreground lens and background source light. High uncertainties in the best-fit parameters for the models due to the limits of optical resolution in ground-based observatories and the small sample size can be improved with future study. We give broadly applicable recommendations for future efforts, and with proper application this approach could yield measurements in the quantities needed for robust statistical inference.

\end{abstract}




\section{Introduction}
\label{s:intro}

Strong gravitational lensing is an essential probe of galaxy structure, enabling mass measurements in the center-most regions of the foreground lensing galaxy without assumptions about stellar populations. Numerous studies have shown that lensing galaxies are, in every other respect, indistinguishable from other galaxies in the observed mass range; therefore, their study offers insight into the larger global population of galaxies at similar mass and redshift \citep{Auger09}. Complementary to kinematic and stellar population synthesis (SPS) measurements, strong lensing allows the decoupling of internal mass components (dark and baryonic) and accurate central stellar population mass-to-light ratios \citep{slacs9,Hopkins18}. 

A fundamental issue in astronomy is relating the predicted dark matter halo mass function to the observed galaxy mass function. The masses of dark matter halos are not well-constrained by the amount of visible matter in their constituent galaxies. Few single galaxies corresponding to the lowest and highest halo masses are found, due to feedback processes stopping star-formation \citep{Behroozi10,Behroozi20}. 
The galaxy stellar-to-halo mass relation (SHMR) represents a fundamental barometer for accretion
and feedback processes in galaxy formation.
Subhalo abundance matching (SHAM) assigns a galaxy with a specific stellar mass to a specific subhalo but does not consider (i) the enveloping host halo mass or (ii) whether the galaxy is a central or a satellite; thus it suffers from \textit{assembly bias} \citep{Zentner14,Chaves-Montero16}.

Assembly bias is a secondary halo property that is related to the clustering strength of haloes \citep{Matthee17a,Zehavi18}, where the clustering of dark matter halos depends on their mass and formation epoch. 
Investigations with cosmological simulations have revealed that dark matter halo concentration, formation time, and environment all play a role in the relation between the central galaxy's stellar mass and the mass of the dark matter halo it occupies. Assembly bias appears to be mostly independent of the cosmological parameters assumed \citep{Contreras21}.
At high masses typical of lensing elliptical galaxies, inefficiency in the stellar occupancy of dark matter haloes is ascribed to the effects of AGNs \citep{Somerville15}. Weak lensing studies \citep{Velander14,Mandelbaum13,Lin16} and velocity field studies \citep{McCarthy21,Posti21} appear to show the effects of assembly bias on the scales of galaxy populations \citep{Cui21}; assembly bias is especially noticeable at $\sim 1$Mpc scales \citep{Hearin16}: i.e. groups of galaxies.  While well-explored with hydrodynamical simulations \citep[e.g.][]{Hearin15,Hearin16,Matthee17a,Artale18,Zehavi18,Zehavi19,Contreras19}, observational studies have thus far been limited by the need to average over large numbers of similar-mass central elliptical galaxies to obtain a weak lensing or velocity signal.

With strong lensing, one has the opportunity to directly measure stellar and halo masses in elliptical galaxies. Relations between the environment and internal structure of elliptical galaxies have been explored using SLACS lenses (Sloan Lens ACS) \citep{slacs1} by \cite{Treu10}. They find that the SLACS lenses are slightly biased toward overdense environments (12 of 70 are associated with known groups or clusters), which is consistent with the expectation for the most massive of elliptical galaxies. They find this result to be unbiased when compared to similar massive galaxies from SDSS, again showing lens galaxies to be representative of the overall elliptical galaxy population. They find the contribution of the external environment to have little effect on the local potential (except in extreme overdensities) and the internal structure of lens galaxies. SLACS and other lens studies have been conducted using detailed observations of individual lenses with HST-quality data. The application of lens modeling methods to larger wide-field surveys offers an alternative avenue with advantages for conducting experiments relating galaxy properties to environment and group properties. This motivates the need for exploring lens modeling methods in the context of large surveys.

In this paper we explore what can be done with ground-based imaging and spectroscopy to model lens candidates after they have been identified in imaging surveys. We discuss strategies for ensuring quality control at each stage while extracting meaningful measurements from ground-based data. Using Galaxy and Mass Assembly (GAMA) survey single-aperture spectroscopy, we explore the utility of automated redshift determination as a tool for identifying the background-source redshifts of strong lenses by applying this technique to lens candidates that were identified in the ground-based imaging of the Kilo-Degree Survey (KiDS) using machine learning techniques \citep{Petrillo19,Li20}. With GAMA spectroscopic redshifts and other measurements in conjuction with KiDS cutout images, we construct lens models utilizing an automated lens modeling program called {\sc PyAutoLens}.

Our paper is organized as follows:
Section \ref{sect:data} describes the KiDS and GAMA data used, as well as the parent samples used in our selection.
Section \ref{sect:autoz-z2} describes how background-source redshifts are identified in single-aperture spectra from GAMA {\sc Autoz} catalogs to create a subsample for modeling.
Section \ref{sect:model} describes the {\sc PyAutoLens} software and the lens modeling methods we used to perform our analysis.
Section \ref{sect:quality_scoring} outlines the assessment of quality of the models and redshift determinations.
Section \ref{sect:model_results} presents results for the four highest-quality models.
Section \ref{sect:autoz_quality} discusses some challenges that our prescription of second-redshift determination introduced, as well as recommendations for improving that method.
Section \ref{sect:discussion} discusses galaxy environment and potential applications of a refined method to future studies.
Section \ref{s:conclusions} lists our conclusions. 
Throughout this paper we adopt a \cite{Planck-Collaboration15} cosmology ($H_0 = 67.7$ km/s/Mpc, $\Omega_m = 0.307$). 

\section{Data}
\label{sect:data}

\subsection{GAMA Spectroscopy and AUTOZ Redshifts}

Galaxy and Mass Assembly \citep[GAMA,][]{Driver09, Driver11,Liske15} is a multi-wavelength survey built around a deep and highly complete redshift survey of five fields with the Anglo-Australian Telescope.  
GAMA has three major advantages over SDSS in the identification of blended spectra:  (i) the spectroscopic limiting depth is 2 magnitudes deeper ($m_r<19.8$ mag compared with SDSS main survey depth $m_r < 17.7$ \citep{Eisenstein01}\footnote{The spectroscopic luminous red galaxy (LRG) sample used to select SLACS lenses is limited to $m_r < 19.5$ thanks to the 4000\AA{} break.}), (ii) the completeness is close to 98\% \citep{Liske15}, and (iii) the {\sc Autoz} redshift algorithm can identify spectral template matches with signal from two different redshifts \citep{Baldry14}. 
These properties and the overlap of GAMA and KiDS fields make these two surveys exceptionally well-suited to provide the data required for our study of lens modeling.

The {\sc Autoz} \citep{Baldry14} cross-correlation redshift software has been uniformly applied to the GAMA \citep{Liske15} spectroscopic data, resulting in a public database that can be found in GAMA-DR3 {\sc AATSpecAutozAll v27} (hereafter {\sc Autoz} catalog) and {\sc SpecAll v27} tables (\url{http://www.gama-survey.org/dr3/}). The {\sc Autoz} algorithm outputs four flux-weighted cross-correlation peaks (denoted $\sigma$) of redshift matches to template spectra of emission-line and passive galaxies (denoted ELG and PG respectively) from SDSS-DR5. $\sigma_1$ corresponds to the highest cross-correlation or "best-fit" redshift match, $\sigma_2$ the second-best, etc. These matches have proven to be highly successful and are the base redshift measurement for GAMA objects. GAMA-DR3 also compiled SDSS-BOSS spectra for overlapping targets that are included in table {\sc SpecAll}, but these spectra did not utilize {\sc Autoz} for redshift determination.

\cite{Holwerda15} analyzed {\sc Autoz} catalog cross-correlation outputs and identified 104 strong lensing candidates from their blended spectra, all of which showed a passive galaxy (PG) with an emission line galaxy (ELG) at higher redshift between cross-correlation $\sigma_1$ and $\sigma_2$. This identification selected candidates from a two-dimensional parameter space defined by the second cross-correlation peak $\sigma_2$ and the parameter $R$, which describes the significance of $\sigma_2$ compared to the following "poorer" matches: 

\begin{equation}\label{equation:R}
    R = \frac{\sigma_2}{\sqrt{\frac{\sigma_3^2}{2} + \frac{\sigma_4^2}{2}}}
\end{equation}

\noindent
Candidates with second cross-correlation peak $\sigma_2\geq4.5$ and $R\geq1.85$ were considered likely candidates for strong lensing. \cite{Knabel20} further analyzed and made a cleaner selection of 47 candidates.

The completeness of GAMA allows detailed environment measures including population density and separation \citep{Brough11,Alpaslan14,Alpaslan15}, and the GAMA team internal {\sc GroupFinding} catalogs include the total mass and placement of the galaxies in an identified group via a friends of friends algorithm \citep{Robotham11}. In fact, GAMA was conceived to probe the effects of group environment on galaxy properties. 
In this context we describe galaxies either as "group member" galaxies or as those not in galaxy groups, which we designate as "isolated galaxies". Stellar masses are taken from the GAMA-DR3 {\sc StellarMassesLambdar v20} catalog \citep{Taylor16}.

\subsection{Kilo-Degree Survey (KiDS) and Machine Learning Strong Lens Samples}

The Kilo-Degree Survey \citep[KiDS,][]{de-Jong13,de-Jong15,de-Jong17,Kuijken19} is a VLT Survey Telescope (VST) program of medium-deep imaging in SDSS-\textit{ugri} filters primarily to identify weak lensing. The deep imaging, high resolution (0.65 arcsec seeing in SDSS r-band), and wide sky-coverage (1350 deg$^2$) also make this survey ideal for efforts to identify strong lens candidates from imaging. Image-based deep learning efforts have been the most promising of recent developments in automated lens-finding algorithms. Their efficiency and versatility make them ideal for astronomical classification problems involving large datasets, including the detection of strong gravitational lenses (e.g. in Subaru Hyper-Supreme Cam \citep{Speagle19}, DECAM \citep{Huang20}, and Dark Energy Survey data \citep{Jacobs19}). \cite{Petrillo17} developed a machine learning method to identify strong lenses in KiDS using a convolutional neural network (CNN) with artificially-constructed lens images as the training set. Training and target catalogs were intentionally constructed utilizing SDSS-LRG \citep{Petrillo17} color-magnitude selection cuts to return the largest of known strong lenses that result in the most readily identifiable lens features (Einstein radii close to and greater than 1 arcsecond). The result is the Lenses in KiDS sample \cite[LinKS,][]{Petrillo18,Petrillo19}\footnote{\url{https://www.astro.rug.nl/lensesinkids/}} of some 1300 strong lensing candidates, 421 of which overlap with the equatorial regions of the Galaxy and Mass Assembly (GAMA) survey. \cite{Knabel20} compared data from LinKS objects with the {\sc Autoz} spectroscopic identifications in GAMA \citep{Holwerda15} as well as with KiDS-GalaxyZoo \citep[][Kelvin et al. {\em in prep.}]{Holwerda19} citizen science identifications in the overlapping equatorial fields. A disparity between the candidate samples in terms of stellar mass and redshift is attributed to selection effects. Of the subsample of 421 LinKS candidates in GAMA (hereafter referred to as "LinKS" or "LinKS in GAMA" sample) there was no overlap with the subsample of 47 GAMA spectroscopic candidates. \cite{Knabel20} further subselected 47 LinKS candidates to represent the highest quality of the sample (hereafter referred to as "LinKS from Knabel-2020" sample).

\cite{Li20} followed a slightly modified approach from the lens-search prescription utilized by the LinKS team to search for strong lens candidates in KiDS-DR4. They included several more LRGs and applied their CNN to a sample of "bright galaxies" (BG) that did not undergo LRG color-magnitude cuts. Their search returned some LinKS candidates and resulted in 286 new candidates within the KiDS survey, 48 of which were identified in the GAMA equatorial regions. 39 of those have matches in the {\sc StellarMassesLambdar} mass catalog, and there are no overlaps between this sample and that of GAMA spectroscopy or GalaxyZoo. This candidate sample, which we will refer to as "Li-BG", shares essentially the same parameter space as LinKS candidates, even with the exclusion of the LRG selection \citep{Knabel20}. This is not necessarily surprising considering \cite{Petrillo19} report no significant advantage to the inclusion of color images in the CNNs, as the networks appear to focus more on morphological features and brightness than color separation. Still, the extension beyond the typical red elliptical galaxy as candidate objects suggests the potential for more variability in candidate characteristics.

\section{AUTOZ Second Redshift Selection and Quality Control} 
\label{sect:autoz-z2}
We examine the {\sc Autoz} cross-correlation values for each LinKS and Li-BG lens candidate. Each candidate has been matched to the closest GAMA object by right-ascension and declination within a positional tolerance of 2 arcseconds. Not every object in the equatorial fields is featured in the {\sc Autoz} catalog; these candidates are removed from this study. Some of the objects feature duplicated entries, some of which have conflicting {\sc Autoz} outputs, which we retain for examination and selection.

\subsection{Selection Criteria}\label{sect:selection_criteria}

\begin{figure}
    \centering
    \includegraphics[width=\columnwidth]{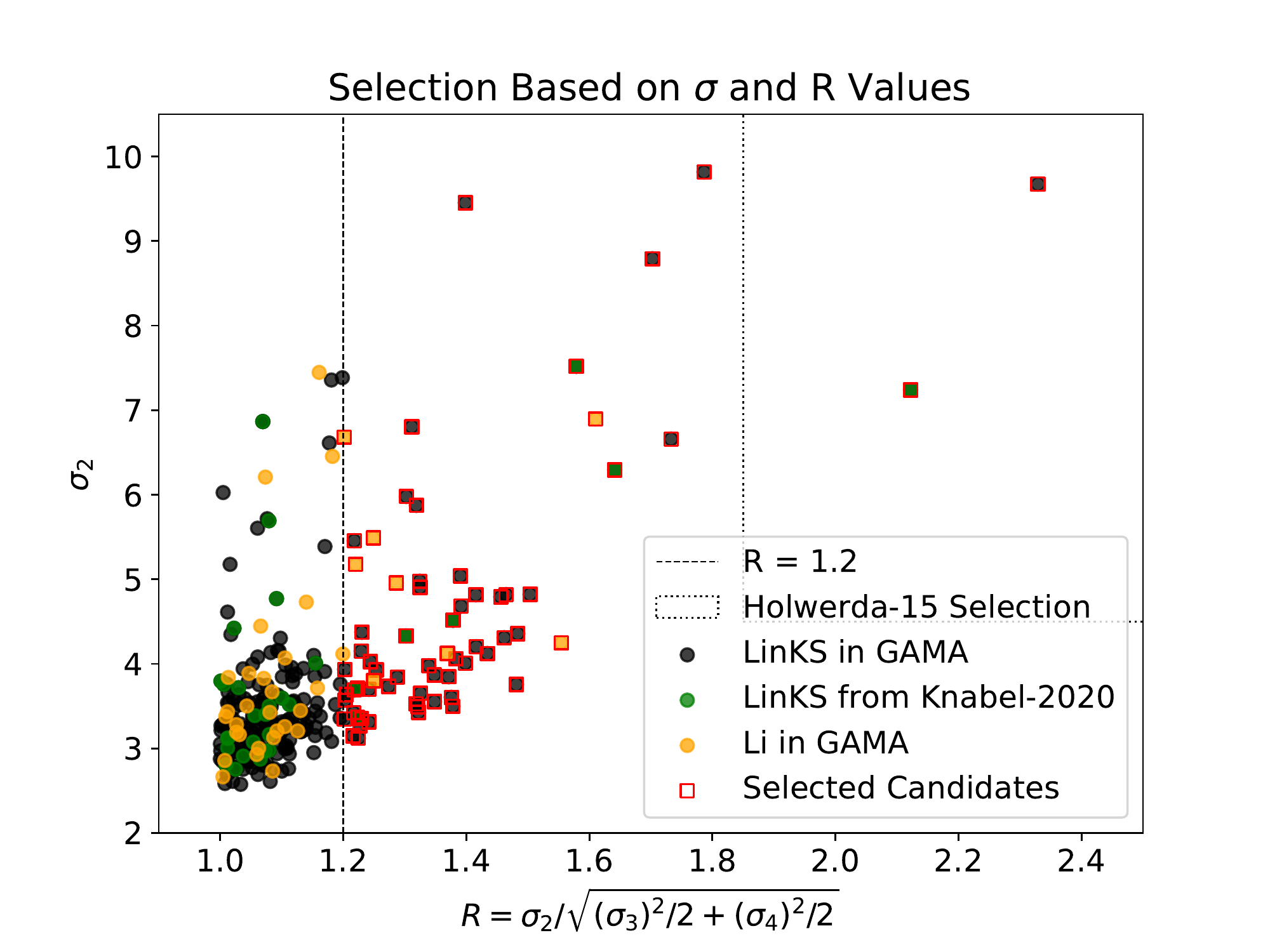}
    \caption{Initial selection of candidates with {\sc Autoz} second-redshift determinations. The y-axis shows the $\sigma_2$ second cross-correlation peak, with higher values indicating a stronger match to the second galaxy template. The x-axis is the parameter \textit{R}, given by Equation \ref{equation:R}. Black markers indicate 348 LinKS {\sc Autoz} entries (300 unique candidates), 51 of which are overplotted with green markers to indicate {\sc Autoz} entres of the 47 unique LinKS candidates as selected in \citealp{Knabel20}. Orange markers indicate the 53 {\sc Autoz} entries for the 32 unique Li-BG candidates. The dashed vertical line shows $R=1.2$, and the dotted box encloses the area of parameter space used by \citealp{Holwerda15}. Red squares surround 59 LinKS and 8 Li-BG entries that satisfy $R\geq1.2$ and are followed with additional selection criteria. See Section \ref{sect:selection_criteria}. }
    \label{fig:sigma_selection}
\end{figure}

Since the candidates that remain have already been identified and vetted through machine learning methods, we adopt a more lenient selection criterion from the same $\sigma_2-R$ parameter space as that utilized in \cite{Holwerda15}. From a first look at the data, we select candidates with $R\geq1.2$. This is sufficient for a first selection and for characterizing the output of the {\sc Autoz} algorithm from already positively-identified candidates. We show the selection in Figure \ref{fig:sigma_selection}.

\begin{figure}
    \centering
    \includegraphics[width=\columnwidth]{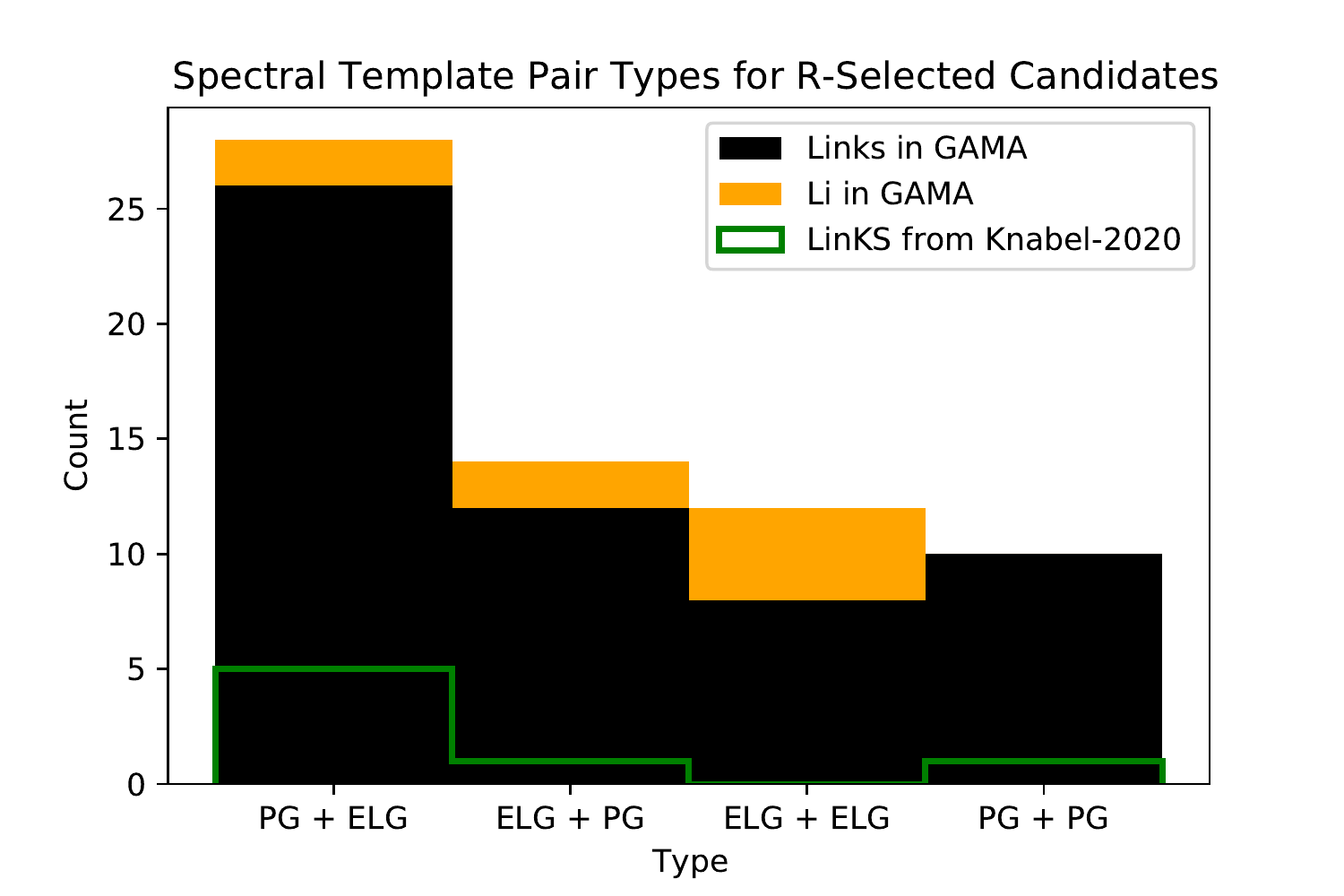}
    \caption{{\sc Autoz} galaxy template combinations of candidate subsamples, listed as foreground+background. Black and orange refer to LinKS and Li-BG candidates respectively and are stacked to show total counts of each template combination. The green outline shows the subset of LinKS  candidates that were examined in \citealp{Knabel20}. As expected, the majority of lens foreground objects match to PG templates, while the majority of background objects match to ELG templates.}
    \label{fig:template_types}
\end{figure}

From the 67 {\sc Autoz} entries that pass the $R$ selection, we remove those with stellar template matches (i.e. not a galaxy spectrum) and retain all those with galaxy-galaxy template match configurations regardless of the galaxy type. The distribution of foreground+background (lens+source) template type (PG+ELG, ELG+PG, ELG+ELG, PG+PG) is shown in the histogram of Figure \ref{fig:template_types}. Note that the majority of lens foreground objects match to passive galaxy templates, with the majority of background objects matching to emission line galaxy templates. This is expected. Massive elliptical galaxies tend to be the most readily observable strong lensing foreground objects, and bright emission lines from the background source are the most easily detected behind a passive galaxy continuum. This selection bias is further enhanced by the fact that the parent LinKS and Li-BG samples were both identified by CNNs trained with large elliptical galaxies. Configurations with foreground lens emission line galaxies are possible, though much more difficult to detect. The reasons are: (i) emission line galaxies are typically lower in mass, so the lensing is less pronounced, (ii) lensed background sources are often bright emission line galaxies, so the blue light of each can blur together, and (iii) emission line galaxies can include complex morphologies that can be mistaken for lens features. The majority of Li-BG candidates in the {\sc Autoz} catalog were matched to emission line galaxies in the foreground, but further selection and assessment of the spectra showed this to be a false trend. As in \cite{Knabel20}, we select candidates with background source redshifts that are reasonably far away from the foreground lens redshifts, as well as those whose {\sc Autoz} foreground redshifts are greater than 0.05. {\sc Autoz} estimates the probability of success of the primary redshift match, and one candidate is removed due to its very low probability. 

Described in more detail and discussed in the context of \cite{Knabel20} in Appendix \ref{s:kids:2zs}, our selection results in 42 lens candidates (39 from LinKS and 3 from Li-BG) with second redshift matches, which constitute the initial {\sc Autoz} sample. In later work, it may be worth exploring machine learning algorithms to find better use of the parameter space for classification than our naive selection criteria.

\subsection{AUTOZ Selection Quality Control} \label{sec:autoz_cases}

An initial modeling and analysis strategy utilizing the {\sc Autoz} sample of 42 candidates and {\sc PyAutoLens} revealed the need for further quality control. This assessment addresses the validity of our application of {\sc Autoz} output parameters to select second-redshift determinations for the use in strong lens modeling. We examined the 42 candidate spectra to understand the evidence for a second redshift match within each. We superimposed upon the candidate spectra important emission and absorption features redshifted by the lens and source redshifts determined by {\sc Autoz}. For ELG template matches, we inspected $H{\beta}$, [O II]$\lambda3727$, and [O III]$\lambda\lambda$4959,5007 emission lines. For PG template matches, we looked at Calcium H and K, $H{\beta}$, Mg, and Na absorption lines. This illuminated some specific cases that pointed to dubious redshift matches. We identify four such cases:

\subsubsection{When there are Overlapping Emission or Absorption Lines...}

{\sc Autoz} template matches utilize strong absorption or emission features to identify passive and star-forming galaxies. We discovered a significant failure condition of our method that occurs when the second redshift match is identified by an emission or absorption line that is also attributed to the foreground lens galaxy redshift. This is particularly obvious in cases where {\sc Autoz} determined ELG+ELG configurations (i.e. both redshifts are identified by emission lines), in which many of the higher-redshift matches were determined by an [O II]$\lambda3727$ line that overlapped with the [O III]$\lambda\lambda$4959,5007 lines of the foreground lens galaxy. For configurations where both the lens and source are the same template type (ie. PG+PG or ELG+ELG), overlapping line features usually indicate a poor second-redshift determination.

However, emission and absorption features are present in the spectra of both PGs and ELGs. For example, \cite{Treu02} found that $\sim10\%$ of elliptical galaxies in the intermediate redshift range where our lens candidates lie have strong [O II]$\lambda3727$ emission; in fact, the emission of [O II]$\lambda3727$ is detectable in the spectra 14 of the passive galaxy matches in our sample, and absorption of Calcium H and K lines (as well as some Na and Mg) are identifiable in 15 of the ELG galaxy spectrum matches. In most cases, the overlap of lines draws suspicion of the background source spectrum match, not the foreground lens match, which is typically the primary template match 
($\sigma_1$). In our sample, emission line overlaps occur exclusively between the background source [O II]$\lambda3727$ and one of the [O III]$\lambda\lambda$4959,5007 couplet of the foreground lens passive galaxy. Absorption line overlaps are all between source H or K lines and lens H$\beta$, Na, or Mg lines.



\subsubsection{When the Lens is described as an Emission Line Galaxy...}

As shown in Figure \ref{fig:template_types}, several of the template matches selected by the strategy described in Section \ref{sect:selection_criteria} are configurations with a foreground lens emission line galaxy (ELG+ELG or ELG+PG). We reiterate that there is no physical reason to mistrust these redshift matches on this fact alone. In fact, given that several of the ELG matches have strong absorption features at the same redshift, these may very well be elliptical galaxies with strong oxygen emission lines. However, upon examination of candidate spectra and the quality of initial models for these ELG+ lens configurations, we found that several of these candidates included dubious template matches and were removed.

The overlapping ELG+ELG emission lines have been discussed. In addition, some of the spectra of ELG+ configurations included key emission line features that would have been observed in wavelength ranges of high noise. Many of the GAMA spectra have their most significant noise at the extremes of the wavelength range, e.g. at wavelengths shorter than $\sim$4500\AA. For lenses at redshifts lower than $\sim$0.2, the [O II]$\lambda3727$\AA emission line lies in this region, which removes an important identifying emission line feature from consideration.

\subsubsection{When Source Emission Lines are Redshifted Beyond Observed Wavelength Range...}

The cases where a background source emission line overlaps with a foreground lens emission line often correlate with another case: some of the background source emission lines have redshifted to wavelengths beyond the observed wavelength range of the spectroscopic survey. For our purposes, this translates to upper limits of background source redshifts beyond which the emission line will not be present in the spectrum.  For GAMA, which has an upper limit of 8850{\AA}, the emission lines we have discussed begin to disappear around z$\sim$0.77. SDSS-BOSS has an extended upper limit of wavelength to 10400 {\AA} ($\sim$J-band), which corresponds to upper redshift limits of around z$\sim$1 where we begin to see missing emission lines. GAMA-DR3 {\sc SpecAll} table contains SDSS-BOSS spectra for several of the candidates, so we can look to these spectra for evidence of lines that have redshifted beyond the GAMA upper limit. The {\sc Autoz} catalog includes only GAMA spectra, so the outputs we use for selection would not benefit from the extended range of the SDSS-BOSS spectra.

\subsubsection{When Primary Redshift is Background Source...}

The primary redshift match corresponding to $\sigma_1$ is typically (but not always) the foreground lens galaxy. The background source redshift is the primary redshift match for 10 of the 42 {\sc Autoz} sample candidates. The results of the {\sc Autoz} algorithm are largely flux-weighted, and an especially bright background source (e.g. a strong emission line) could be interpreted by the algorithm as the primary redshift solution instead of the lower-redshift lens object. However, 8 of these 10 are ELG+PG template configurations, where a PG template gives the primary redshift solution at higher redshift. The case of a bright continuum at higher redshift overshadowing the foreground emission line galaxy is unlikely.

All candidates in the {\sc Autoz} sample were modeled and examined in the manner described in Sections \ref{sect:model}-\ref{sect:model_results} before these four cases were identified. In Section \ref{sect:autoz_quality} we discuss the results of modeling and assessment in the context of the cases described in this section and recommend alterations to the initial selection scheme. We note that a poor redshift match does not remove/negate the validity of the lens identification, nor does it question the accuracy of the uniform application of {\sc Autoz} to GAMA spectroscopic targets. These additional selection decisions were instituted following critical assessment of problems with our initial strategy that required time with human eyes on the spectra. With reasonable background source redshift determinations, modeling of the imaging data can yield meaningful physical measurements.

\section{PyAutoLens}
\label{sect:model}


We use the open source lens modeling software {\sc PyAutoLens} \citep{pyautolens}\footnote{\url{https://github.com/Jammy2211/PyAutoLens}} to perform our analysis. The software is described in \citet{Nightingale2018} and \citet{pyautolens}, building on the works of \citet{Warren2003}, \citet{Suyu2006} and \citet{Nightingale2015}. We refer readers to these works for a full description of our approach to lens modeling. Section \ref{sect:pyautolens_description} broadly describes the method as we apply it here, and a more technical description of the specifics of the implementation is given in Appendices \ref{s:prepdata} and \ref{sect:pipeline}.

\subsection{Lens Modeling with PyAutoLens}\label{sect:pyautolens_description}

{\sc PyAutoLens} models the foreground lens galaxy's light and mass as well as the background source galaxy's light simultaneously. First, {\sc PyAutoLens} assumes a profile for the foreground lens's light (e.g. a S\'ersic profile), producing a model image of the lens galaxy. A mass model then ray-traces a grid of image-pixels from the image-plane to the source-plane, with the source's light evaluated on this deflected grid via another light profile. This creates an image of the lensed source, which is added to the lens galaxy image to create an overall model image of the strong lens. This image is convolved with the instrument PSF and compared to the data to evaluate the residuals and likelihood of that lens model. To fit a lens model to imaging data, {\sc PyAutoLens} searches an N-dimensional parameter space so as to minimize the residuals (and therefore maximize the likelihood) between the model image and the observed image. The lens models fitted in this work consist of $N=7-14$ parameters, corresponding to the parameters of the light and mass profiles that represent the lens and source galaxies. To sample parameter space, we use the nested sampling algorithm {\sc Dynesty} \citep{Speagle19}, and we detail its specific implementation below.

The parameter spaces of a strong lens model are challenging to sample, and local maxima and unphyscial lens models are often inferred. Automating the model-fitting procedure is therefore difficult, and {\sc PyAutoLens} approaches automation via a technique called non-linear search chaining. Here, a sequence of {\sc Dynesty} model-fits are performed that fit lens models of gradually increasing complexity, whereby the results of the initial searches are used to inform the search of more complex parameter spaces in the later searches. Through experimentation, we have designed a pipeline composed of a chain of three {\sc Dynesty} searches that we use as a template for fitting each lens. Our three-step pipeline consists of three sequential model-fits:

\begin{enumerate}
    \item \textit{Search 1 - Lens Light}: models only the foreground lens elliptical light profile.
    \item \textit{Search 2 - Lens Mass and Source Light}: focuses on the background source light profile and lensing deflections.
    \item \textit{Search 3 - Combined Lens and Source Models}: models each component in the system for parameter inference.
\end{enumerate}

Between each search, various aspects of the fit can be altered (e.g. a mask applied to the data can be customized to show only the specific features of interest to each fit). This offers a more efficient lens modeling procedure overall, as the parameter spaces of reduced complexity are sampled faster. 

Search chaining uses a technique called "prior passing" to initialize the regions of parameter space that are searched later in the chain. Here, the models inferred in earlier non-linear searches initialize the priors of the more complex models fitted by the searches later on. This ensures the non-linear search samples only the higher likelihood regions of parameter space (see \citet{Nightingale2018}) and therefore reduces the probability that a local maximum is inferred. Prior passing sets the prior of each parameter as a Gaussian. The mean is that parameter's previous inferred median PDF value, and the width is a value specific to each lens model and parameter. Prior widths have been carefully chosen to ensure they are broad enough not to omit lens model solutions by trimming valid solutions but sufficiently narrow to ensure the lens model does not inadvertently infer local maxima.

The {\sc Dynesty} nested sampling algorithm \citep{Speagle19} can also balance efficiency in computation with how thoroughly it explores parameter space. Initial model fits require only a rough estimate of the lens model that provides a reasonably approximate fit to the data. These searches therefore use faster {\sc Dynesty} settings that give a less thorough sampling of parameter space. Our final results require accurate and robust parameter estimates with precise and well-quantified errors. By Search 3, the priors are initialized such that a deeper exploration of the parameter space can be performed more efficiently, ensuring that {\sc Dynesty} does not spend considerable time in regions of parameter space that previous searches tell us do not give a physical lens model. Some basic settings that can be varied to affect the performance of the non-linear search are: (i) number of live points, (ii) number of steps of random walks per iteration, (iii) target acceptance fraction for random walks, (iv) Bayesian evidence tolerance, (v) positions threshold, and (vi) sub-grid size. 

The technical details of our modeling method, including data preparation and pipeline, are described more fully for the interested reader in Appendices \ref{s:prepdata} and \ref{sect:pipeline}. The sequence of chained searches and specific parameters that are set via prior passing are listed in Table \ref{table:phases}, and Dynesty settings are tabulated in Table \ref{table:phases_settings}. More details on {\sc PyAutoLens}'s use of {\sc Dynesty} are provided in (Nightingale et al. \textit{in prep}).

\subsection{Physically Motivated Priors}\label{sect:priors}

Where possible, we calculate priors using photometric observations from GAMA-DR3 preferentially over a universally applied "typical" value. For either case, it is important not to fix the parameters too restrictively to values from observations.

\subsection{Effective Radius}

In Search 1, we initialize the effective radius parameter of the foreground lens light profile with a Gaussian centered at the lesser of two possible radii: (i) effective radius determined from photometric observations from GAMA-DR3 {\sc SersicCatSDSS v09} catalog \citep[][]{Kelvin12}, or (ii) the median SLACS lens effective radius and standard deviation from \cite{Auger10} ($7\pm3.3$kpc). We expect the GAMA-DR3 observation to include extended blended light from the source feature, which may result in a higher measured effective radius than would be measured from the foreground galaxy if it were not lensing. In order to assist the search in the task of deblending the lens and source light, we ensure that the prior is not predisposed to unusually large effective radii. Another failure state of early models resulted in unrealistically large source galaxy effective radii. Instead of attributing the extended lens features to lensing of a compact background object (which is most often the case for strong lensing), the model makes up for that extra flux as the physical extent of an extremely large, bright background source galaxy at high redshift. This motivated an upper limit to the effective radius of the source galaxy based on typical disk galaxy properties. We take a rough value of $7.5\pm2.5$ kpc and upper limit of about 15 kpc.

\subsubsection{Lens Mass-to-Light Ratio} 
\label{sect:ml_prior}

Certain critical parameters are not easily approximated with typical observations (and as such are the goal of the search), such as the stellar mass-to-light ratio. This quantity can be inferred from stellar population studies, but one of our goals is to illuminate mass relations without the dependence on these assumptions. We want the model to tell us about the stellar population as opposed to the inverse. We want the algorithm to have the maximum freedom to determine the best combination of stellar and dark mass profiles to account for a gravitational potential that can describe the observed lensing deflections. Our first attempts allowed for a wide uniform prior distribution for the mass-to-light ratio of the stellar light-mass profile. The resulting models showed higher values than expected, some of which were unphysical in the context of predicted $M_*/L$ from stellar population models evolving with age. The population would have to be older than the age of the Universe at the given lens redshift for the model's $M_*/L$ to reconcile with our current models of stellar populations and evolution.
To approach this problem, we impose a minimum $M_*/L$ of 1 $(M/L)_{\odot}$ and a maximum determined as a function of lens redshift. We assume a Salpeter IMF and utilize stellar evolution models from \citep{Bruzual03} (updated 2016) based on the STELIB spectral library. We determine the maximum possible restframe bandpass-specific $M_*/L$ corresponding to a formation time close to the beginning of the Universe. Other libraries (BaSeL and Milessx) give almost identical values. Given a simple stellar population forming from a single starburst at time $t=0$, Salpeter IMF, and solar metallicity (Z = Z$_{\odot}$ = 0.02 , X = 0.7000, Y = 0.2800, [Fe/H] = +0.0932), we examine the evolution of the stellar mass-to-light ratio with population age in r- and g-bands sampled at unequally spaced time steps over 20 Gyr of stellar evolution. In our adopted cosmology, the age of the Universe in the redshift range of our sample (z$\sim$0.07-0.45) is about 9-13 Gyrs. At this late stage in stellar evolution, the stellar mass-to-light ratio varies slowly and is reasonably approximated as a linear relation. On the domain [9, 13 Gyrs], the constraint is a linear relation:

\begin{align}
    M_*/L_r [M_{\odot}/L_{\odot}] < 0.466 t + 0.719 \label{equation:mls_r} \\
    M_*/L_g [M_{\odot}/L_{\odot}] < 0.717 t + 0.380 \label{equation:mls_g} 
\end{align}
where $t$ is the age of the Universe at the lens redshift in the adopted model cosmology.

In order to be implemented as priors in the lens models, these restframe constraints must be k-corrected, calibrated to the flux units in which the observed data is given, and rewritten in the model mass and intensity units. We use SED-calculated k-corrections from GAMA-DR3 {\sc kcorr\_auto\_z00 v05} \citep{Loveday12} for each lensing galaxy to constrain the prior in the observed bandpass. These constraints are converted to angular mass units per eps (electrons per second) with the gain and exposure time of the KiDS observation. These constraints ensure that the model does not attribute mass to a stellar population that is impossible within current stellar evolution models. In cases where the maximum possible $M_*/L$ is fit, the maximum possible stellar mass has also been attributed. In these cases, the model may end up having to compensate with very high amounts of dark matter to account for the lensing potential. 

\subsubsection{NFW Profile Scale Radius}

The scale radius of a dark matter halo is one of the key parameters of the NFW mass density profile for dark matter halos. \cite{slacs4} modeled 22 SLACS lenses with strong and weak lensing constraints and a two-component mass profile consisting of a de Vaucouleurs stellar component and spherical NFW dark matter component. We adopt their resulting mean scale radius of $r_s = 58 \pm 8 h^{-1}$ kpc for our dark matter profile Gaussian prior distributions. These values are converted to arcseconds from angular diameter distances in our assumed cosmology.

\subsection{Choice of Image Bandpass}\label{sect:bandpass}

KiDS observations of each object in different bandpasses are not equal in exposure time, signal-to-noise quality, or PSF. KiDS r-band imaging is the highest quality of the bandpasses, with an exposure time of 1800 s and a mean PSF of 0.65 arcseconds. g-band exposure times are 900 s. For each candidate and for each model search, we select the better of r- or g-band images. Search 1 fits the r-band image because it most clearly shows the foreground lens light. If the lens and source are clearly distinguishable in the r-band, then the same image is used for Searches 2 and 3. However, the g-band image often most clearly shows the lensed features of the background source; in these cases the g-band is preferable for Search 2. Since Search 3 models the entire system, the image that most clearly shows both profiles is used. 

Each search assists the subsequent searches to distinguish between the lens and source light, which is one of the more difficult challenges of modeling lenses from images of the resolution and S/N attainable by ground-based observatories. Two effective first solutions are (i) separating the initial search of foreground lens and background source light profiles by color-band and (ii) masking specific regions of the image. In this case the sacrifice in quality between the r- and g-bands as a consequence of survey design presents an additional challenge to the fitting process as well as in later analysis. {\sc PyAutoLens} uses units of electrons per second, so the effect of the difference in exposure times and S/N between observations with the two bandpasses is minimized. However, measurements taken from models that fit images of the same bandpass are much easier to compare.

\section{Model Quality Assessment and Grading}
\label{sect:quality_scoring}

With models complete, we assess the highest-likelihood models for each candidate. Ideally, a reliable objective figure of merit such as image $\chi^2$ or Bayesian evidence would sufficiently quantify the quality of each lens model fit. Forming robust quantitative goodness-of-fit metrics is currently an open problem in automated lens modeling. \cite{Etherington22} explored this problem using {\sc PyAutoLens} with much higher resolution images and found that none did a particularly satisfactory job. We take the Bayesian evidence to be the reference figure of merit and follow this with a blind visual inspection of the image, fit, and spectrum of each modeled candidate. We inspect the images and spectrum separately in order to isolate some of the failure states that occur for each and have a clear picture of the factors limiting the precision of the models. Three collaborators give a separate score between 0 and 4 for each of the image, fit, and spectrum for each candidate, so that each of the candidates has three scores out of 12 and a total possible score of 36.

\begin{table*}
\begin{center}
\begin{tabular}{l l l | l }
\hline
Grade & Total Score $\geq$ & Spectrum Score $\geq$ & \# Models with Grade \\

\hline
A & 30 & 9 & 2 \\
B & 20 & 6 & 3 \\
C & 16 & 5 & 9 \\
D & 12 & 4 & 5 \\
\hline
\end{tabular}
\end{center}
\caption{Grading scheme based on the total score and spectrum score for each candidate as described in Section \ref{sect:quality_scoring}. We give greater weight to spectrum score because the quality of the {\sc Autoz} redshift determination is essential to deriving meaningful physical results. All graded models have received no "0" scores for any individual quality by any scorer, ensuring that the final set is clean.}
\label{table:quality_grades}
\end{table*}%

\begin{table*}
    \centering
    \begin{tabular}{|l|l|l|l|l|l|l|l|l|l|l|l|}
    \hline
        GAMA ID & ID & RA & DEC & $z_{\mathrm{lens}}$ & $z_{\mathrm{source}}$ & Type & \textit{Scores}: & Spectrum & Total & Grade & $ln$(evidence) \\ \hline
        323152 & 2967 & 130.546 & 1.643 & 0.353 & 0.722 & PG+ELG && 12 & 33 & A & 7.10 \\ 
        138582 & 2828 & 183.140 & -1.827 & 0.325 & 0.433 & ELG+ELG && 11 & 32 & A & 7.47 \\ \hline
        250289 & 2730 & 214.367 & 1.993 & 0.401 & 0.720 & PG+ELG && 8 & 27 & B & 6.28 \\
        62734 & 539 & 213.562 & -0.242 & 0.274 & 0.597 & PG+ELG && 6 & 26 & B & 4.50 \\ 
        513159 & 2123 & 221.917 & -0.999 & 0.289 & 0.701 & PG+ELG && 7 & 23 & B & 7.59 \\ \hline
        3891172 & 3056 & 139.227 & -1.545 & 0.340 & 0.609 & PG+PG && 5 & 24 & C & 6.43 \\ 
        373093 & 2897 & 139.306 & 1.198 & 0.384 & 0.837 & PG+ELG && 5 & 23 & C & 7.31 \\ 
        559216 & 2507 & 176.116 & -0.619 & 0.250 & 0.714 & PG+ELG && 7 & 19 & C & 7.77 \\ 
        3629152 & 1933 & 135.889 & -0.975 & 0.407 & 0.787 & PG+PG && 5 & 19 & C & 7.36 \\ 
        3896212 & 1483 & 129.806 & -0.830 & 0.382 & 0.848 & PG+PG && 6 & 18 & C & 6.38 \\ 
        342310 & 2163 & 215.081 & 2.171 & 0.380 & 0.693 & PG+ELG && 5 & 18 & C & 5.79 \\ 
        272448 & 2541 & 179.420 & 1.423 & 0.272 & 0.889 & PG+ELG && 5 & 17 & C & 7.07 \\ 
        262874 & 26 & 221.611 & 2.224 & 0.386 & 0.859 & PG+PG && 6 & 16 & C & 6.00 \\
        387244 & 1819 & 135.569 & 2.365 & 0.218 & 0.712 & PG+ELG && 5 & 16 & C & 7.37 \\ \hline
        569641 & BG3 & 219.730 & -0.597 & 0.360 & 0.826 & ELG+ELG && 4 & 25 & D & 7.27 \\ 
        419067 & 1179 & 138.620 & 2.635 & 0.188 & 0.764 & PG+ELG && 4 & 22 & D & * \\ 
        16104 & BG1 & 217.678 & 0.745 & 0.287 & 0.849 & PG+ELG && 4 & 19 & D & 7.08 \\ 
        561058 & 3349 & 182.560 & -0.495 & 0.320 & 0.856 & PG+ELG && 6 & 14 & D & 6.96 \\ 
        262836 & 1953 & 221.405 & 2.314 & 0.144 & 0.418 & ELG+PG && 5 & 13 & D & 7.80 \\ \hline
    \end{tabular}
    \caption{The 19 models with letter grades as selected in Section \ref{sect:quality_scoring}. The other 24 models were considered too poor for consideration here. ID refers to internal the LinKS sample identifier or our labeling of Li-BG candidates that were modeled. Type refers to the configurations of foreground+background galaxy template type. Scores are the sums of scores given by three individual scorers. Grades classify the quality according to the grading scheme shown in Table \ref{table:quality_grades}. $ln$(evidence) is the log of the Bayesian evidence reported by {\sc PyAutoLens}. * G419067 had a negative evidence as a result of high image residuals in the center of the lens light profile.}
    \label{table:graded_models}
\end{table*}

The procedure for assigning quality scores is as follows: The collaborator (the "scorer") is shown via a randomized selection either the spectrum or a set of images (observed and model-fit) of a randomly selected candidate. The spectrum and image set are not shown sequentially in order to keep the scores unbiased by each. The spectrum score is based on the detection of redshifted line features that correspond to both the foreground and background redshifts. Wavelength accuracy, strength, and number of detectable line features are considered, in addition to template type and the presence of overlapping line features. The image and fit are scored simultaneously from the set of four observed and model fit images because the fit score is informed by the image score. The image score is based on two images --- (i) the observed image and (ii) the observed image with the model's lens light S\'ersic profile subtracted. The scorer considers how well the two images appear to show a well-defined structure outside the central foreground lens light-profile that could be reasonably described as a lens feature. For the fit score, the scorer compares the lens-subtracted model image to the lens-subtracted observed image and examines model background source-plane image. The fit score is influenced by the image score; the fit score cannot be higher than the image score +1. This means that a poor image that is fit perfectly should not get a high fit score, and a good image that is fit poorly should reflect the failure of the model to adequately attribute the image features to the lensing of a background source.

\begin{figure}
    \centering
    \includegraphics[width=\columnwidth]{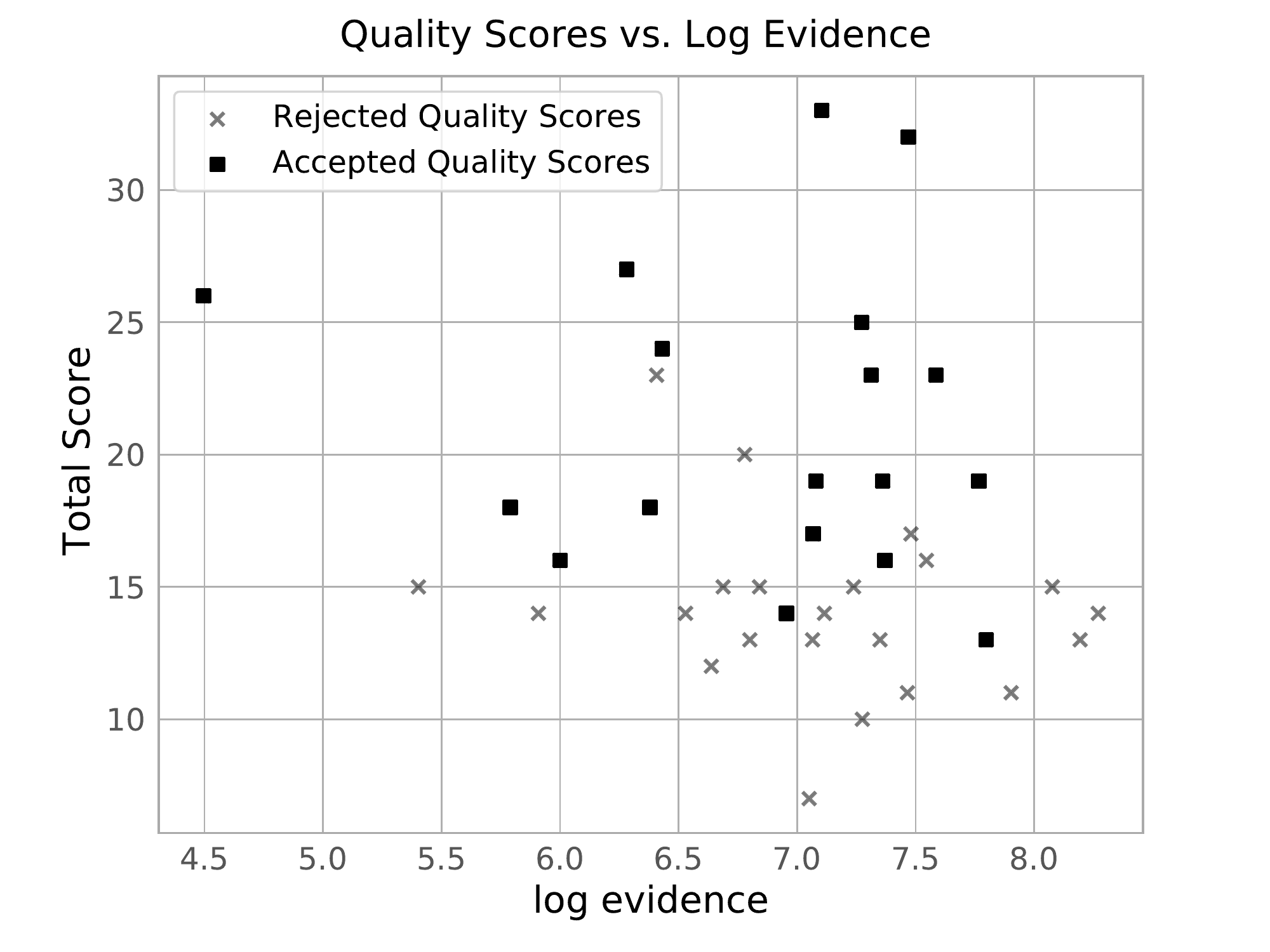}
    \caption{Total score from visual quality inspection vs the natural log of the Bayesian evidence from model fitting. X-markers are rejected based on visual quality inspection. Square markers are accepted. There is very little correlation between the visual inspection results and the objective quality-of-fit metric.}
    \label{fig:quality_score_vs_likelihood}
\end{figure}

Following the scoring exercise, we remove any candidate that received a "0" for any of the image, fit, or spectrum scores by any scorer. This removes catastrophic failures and ensures that the final set is reliable for follow-up analysis. The 19 candidate models that remain are assigned a letter grade of A, B, C, or D according to the structure outlined in Table \ref{table:quality_grades}. There are 2 A, 3 B, 9 C, and 5 D grades in the scored subsample, described in Table \ref{table:graded_models}. 17 of the graded models are candidates from the LinKS subsample. Two of the three candidates from the "Li-BG" sample were modeled to a level of success that justified presentation alongside the others, though both models are given grades of D. One D-grade model (G419067) with a negative likelihood was a result of high image residuals in the very center of the lens light profile. Note the asterisk in the $ln$(evidence) column of Table \ref{table:graded_models}. This model scored well enough for inclusion (spectrum score 4, total score 22) by the blind visual inspection. However, the visual inspection may have removed this candidate with the inclusion of a residual or $\chi^2$ map in addition to the model images. The Bayesian evidence is therefore a prudent first cut of extremely poor models. Otherwise, as shown in Figure \ref{fig:quality_score_vs_likelihood}, we find that the quality of fit determined by careful visual inspection is not correlated to the reported Bayesian evidence.

To the authors' knowledge, none of these lens candidates have been previously confirmed with high-resolution (HST-quality) imaging or spectroscopy. G250289 was identified in HSC as J083726+015639 by \cite{Sonnenfeld19}. Spectrum scores of 6 or better can be considered to be probable spectroscopic evidence for the lens candidate, and the highest spectrum scores for the two A grades can be considered spectroscopic confirmations. No additional extensive efforts were made to identify another possible background source redshift if the one determined by {\sc Autoz} was deemed unreliable. All scores can be considered useful follow-up evidence for the quality of the candidates, in that the success of a model lends additional confidence to the identifications. However, \cite{Petrillo18} and \cite{Li20} have already provided extensive studies of the quality of their lens identifications, and our image modeling is conducted on the same observations as their analysis. Therefore, any poor model performance here does not contradict a positive identification.

\section{Model Results}\label{sect:model_results}

\subsection{Extracting Best-Fit Parameters}

For each lens model, the parameter space is sampled over tens of thousands of iterations, estimating the log likelihood for each sample fit and constructing a probability density function (PDF) for each free parameter listed in Table \ref{table:phases} of the Appendix. Mass and light profiles are fully described by the model-fit parameters. The Einstein radius, total lensing "Einstein" mass, mass fractions, luminosity, and mass-to-light ratios are calculated for each sample from the model grid and integrated over the angular area enclosed by the Einstein radius.

Parameters involving the luminosity require k-corrections to restframe (see Section \ref{sect:ml_prior}). Three of the four models used the g-band image for Search 3, so they are easy to compare. Special attention should be given to G250289, which was instead modeled from its r-band image. In attempting to give the models the best chance to succeed, we were inconsistent in the choice of bandpass for modeling (see Section \ref{sect:bandpass}). Luminosity and mass-to-light ratios for this model are corrected to the g-band after r-band restframe k-correction. This is done by multiplying (or dividing) by an additional factor
$10^{-0.4(g-r)}$,
where $(g-r)\sim0.285$ is the color difference calculated by integrating the product of each bandpass response function and a template elliptical galaxy spectrum from \cite{Kinney96} over the bandpass range. This has the effect of scaling luminosity down and $M/L$ up. Given our goal here, which is to explore the methods, this is sufficient for characterizing the differences between models in a consistent parameter space. However, future efforts that intend to approach these measurements more rigorously should approach the initial modeling with more consistency.

\begin{table*} 
    \centering
    \begin{tabular}{|l|l|l|l|l|l|l|l|l|l|l|l|l|l|l|}
    \hline

        GAMA ID & $z_{\mathrm{lens}}$ & $z_{\mathrm{source}}$ & Type & $M_*/M_{\odot}$ & $M_E/M_{\odot}$ & $f_{DM}$ & $\theta_E$ & $\theta_{eff}$ & $M_*/L_g$ & $L_g/L_{\odot}$ & Grade 
        \\ \hline
        138582 & 0.325 & 0.433 & ELG+ELG & 9.91e+10 & 8.69e+11 & 0.888 & 1.20 & 1.99 & 7.70 & 1.98e+10 & A \\
        323152 & 0.353 & 0.722 & PG+ELG & 1.31e+11 & 4.65e+11 & 0.717 & 1.27 & 2.78 & 4.98 & 2.69e+10 & A \\
        513159 & 0.289 & 0.701 & PG+ELG & 7.31e+10 & 7.29e+11 & 0.900 & 1.72 & 2.44 &  4.85 & 1.52+10 & B \\
        250289 & 0.401 & 0.720 & PG+ELG & 5.47e+11 & 8.82e+11 & 0.375 & 1.55 & 2.44 & 8.92 (\textit{6.86} r) & 6.11e+10 (\textit{7.95e+10} r) & B \\ \hline
    \end{tabular}
    \caption{Results of 4 highest scoring models. $z_{lens}$ and $z_{source}$ are the redshifts of the foreground deflector and background source. Type refers to the configuration of foreground+background template types. $\theta_E$ is the Einstein radius calculated from the model mass distribution and lensing distances. Remaining quantities are integrated within $\theta_E$. $M_E$ is the total enclosed Einstein mass. $f_{DM}$ is the enclosed dark matter fraction. \textit{L} is the enclosed luminosity in the r-band enclosed. $M_*/L$ is the enclosed stellar mass-to-light ratio. Grade is an evaluation of the quality of the fit to the image according to the scheme outlined in Table \ref{table:quality_grades}.}
    \label{tab:final_four_results}
\end{table*} 

We briefly present best-estimate results from the highest-likelihood model fits for the four highest quality models in Table \ref{tab:final_four_results}, selected primarily by the blind quality scoring described in Section \ref{sect:quality_scoring}. Bayesian evidence reported by {\sc PyAutoLens} and the subjective reasonableness of the inferred quantities are also considered in the selection of this small subsample. One B-grade model, G62734, is not included because its central dark matter content is poorly constrained. This and the other 14 lower-grade models are considered to be worth revisiting but were not successful enough to present alongside the cleaner examples we present here.

\begin{figure*}
    \centering
    \includegraphics[width=\columnwidth]{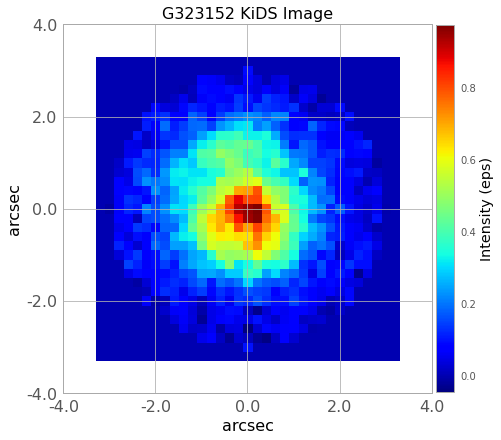}
    \includegraphics[width=\columnwidth]{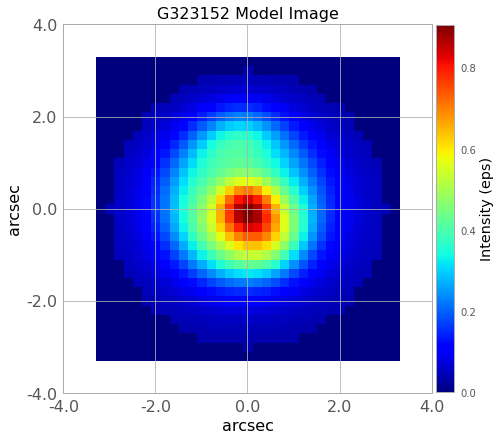}
    \includegraphics[width=\linewidth, trim={3.4cm 0 4.0cm 0}, clip]{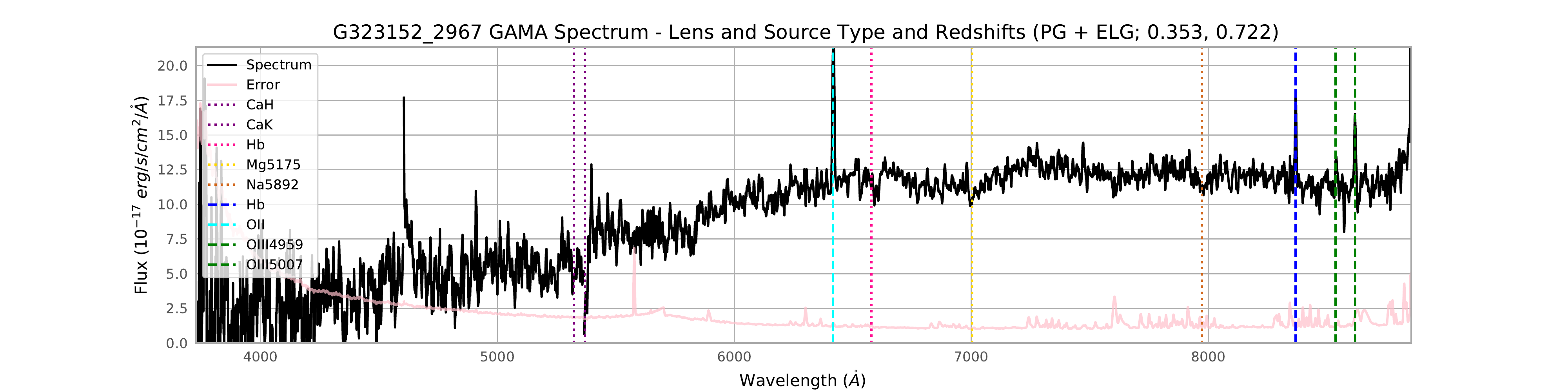}
    \caption{G323152. A-Grade. \textit{Upper left}: The observed image shows an apparent arc feature above the central lens galaxy light-profile. \textit{Upper right}: The model image captures the extra light reasonably, though without the exact shape. This could be a result of internal substructure or the impact of shear along the line of sight, both of which are unaccounted for in the model. \textit{Lower}: The GAMA spectrum shows strong line features at the redshiftsat of 0.353 and 0.722 identified by {\sc Autoz}. Dotted lines identify foreground lens galaxy absorption features (H, K, H$\beta$, Mg, and Na) at $z = $ 0.353, and dashed lines show background source emission features (H$\beta$, [O II], [O III]) at $z = $ 0.722.}
    \label{fig:2967}
\end{figure*}

To discuss inferred quantities, we estimate bivariate PDFs for the quantities using a Gaussian kernel-density estimate from the final 10000 iterations. The best estimate for each inferred quantity listed in Table \ref{tab:final_four_results} is determined at the maximum of one of these bivariate PDFs, where we use uncorrelated values as much as possible. We show the observed image, maximum likelihood model fit, and spectrum for the highest scoring model in Figure \ref{fig:2967}. The other three models listed in Table \ref{tab:final_four_results}, as well as G62734, are shown and discussed in more detail in Figures \ref{fig:2828}-\ref{fig:2123} of Appendix \ref{sect:other_models}.  

We are primarily interested in studying the stellar and dark matter content in the central regions of the lens galaxies. Although the mass and light profiles in the models are inferred to a larger radius, mass measurements via strong lensing are the most precise when considering only mass within the Einstein radius of the galaxy. It is important to note that the Einstein radius is a feature individual to each system, so the quantities are not calculated within a uniform radius from the center of each galaxy. Values for each Einstein radius can be found in Table \ref{tab:final_four_results}. 

\subsection{Comparing Highest-Quality Model Results}

\begin{figure}
    \centering
    \includegraphics[width=\columnwidth]{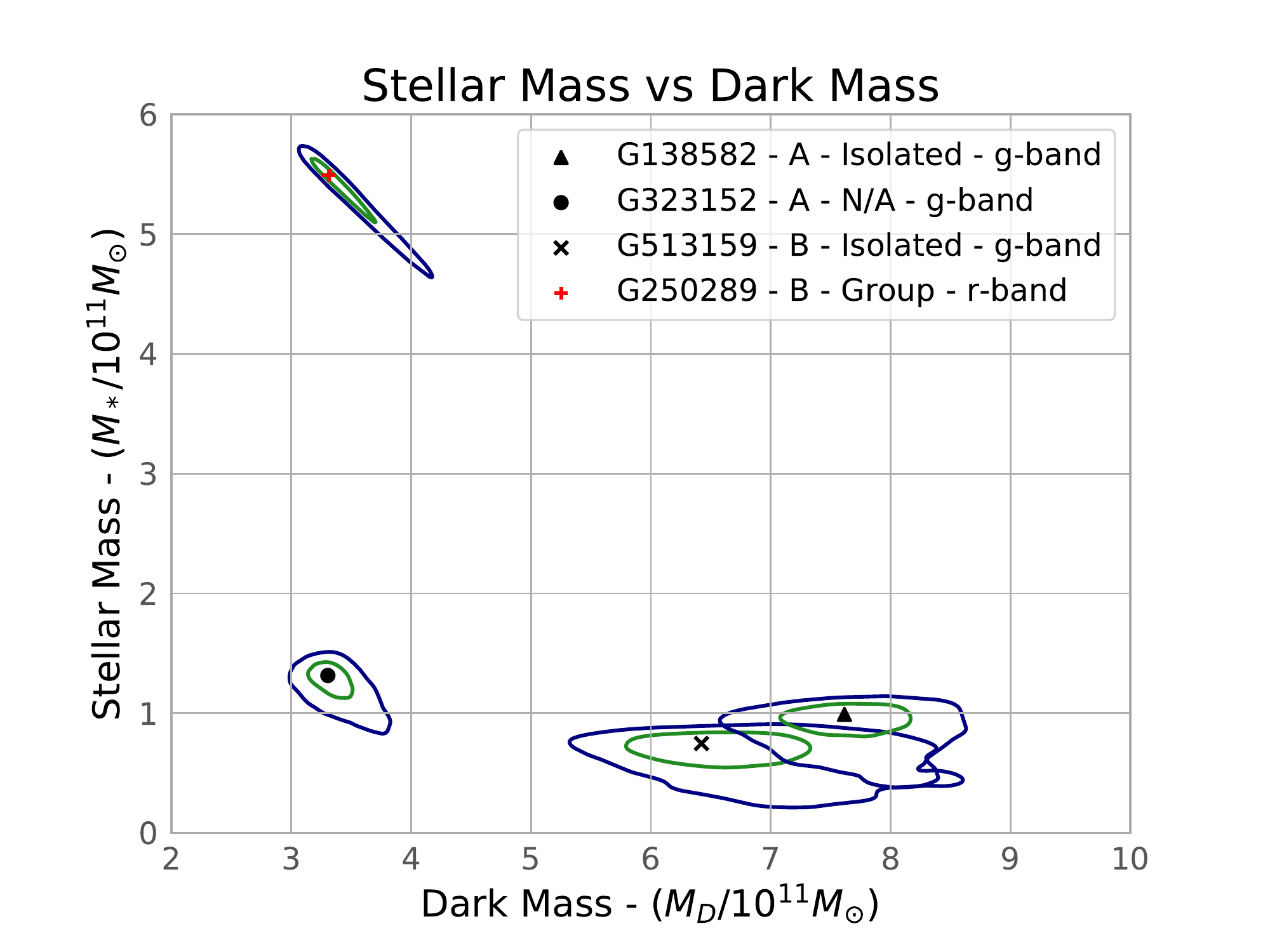}
    \caption{Mass components integrated within the Einstein radius of each of the four best-fit models described in Section \ref{sect:model_results} and Table \ref{tab:final_four_results}. The legend shows the GAMA identifier, quality grade, environment classification, and the SDSS bandpass used for the final model fitting. Green and blue contours about each point enclose $1\sigma$ (39\%) and $2\sigma$ (86\%) of the two-dimensional PDF respectively.}
    \label{fig:sm_dm}
\end{figure}

\begin{figure}
    \centering
    \includegraphics[width=\columnwidth]{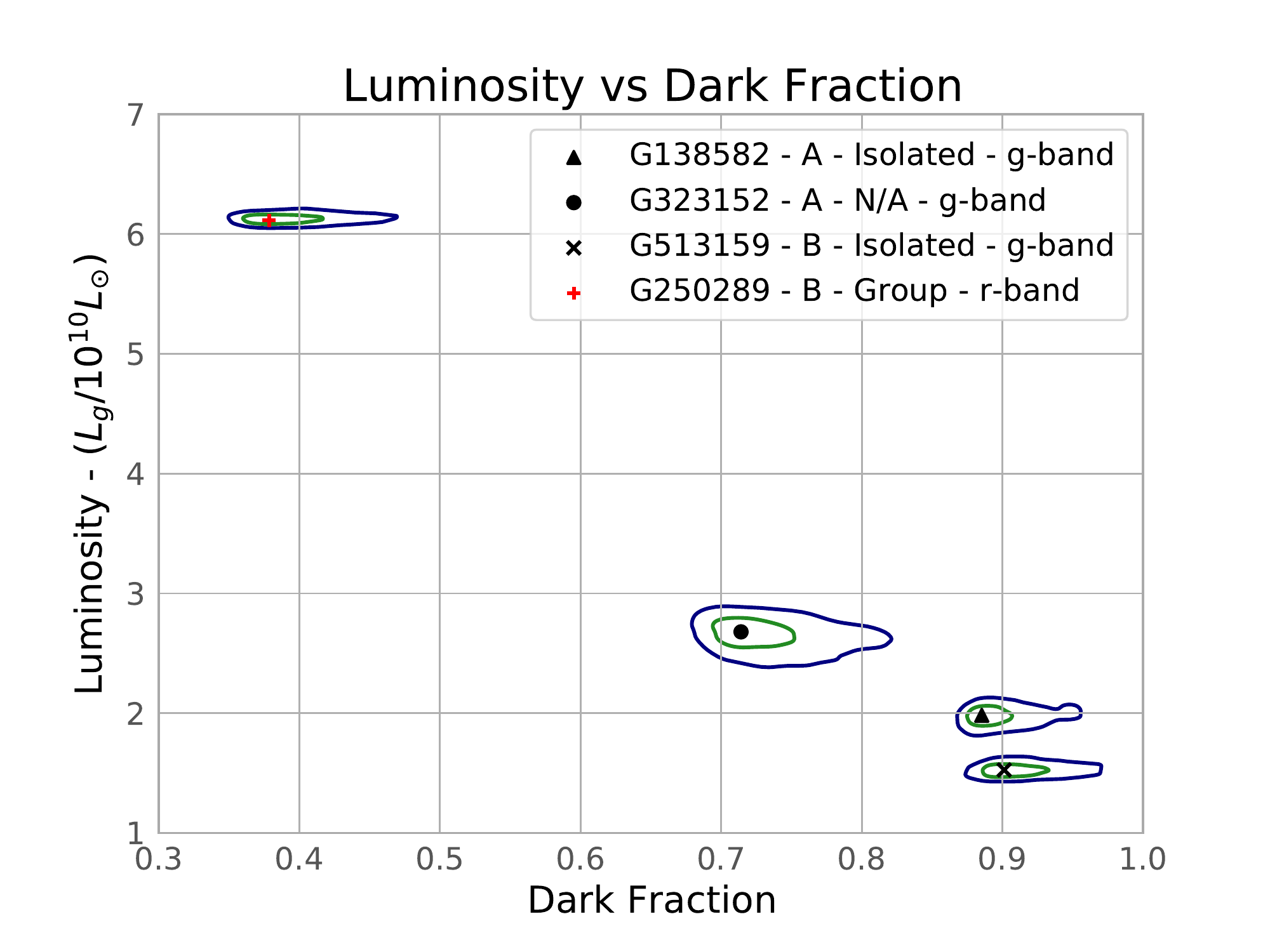}
    \caption{SDSS g-band luminosity and dark matter fraction integrated within the Einstein radius of each of the four best-fit models described in Section \ref{sect:model_results} and Table \ref{tab:final_four_results}. Legend and marker information are the same as in Figure \ref{fig:sm_dm}.}
    \label{fig:lum_fdm}
\end{figure}

\begin{figure}
    \centering
    \includegraphics[width=\columnwidth]{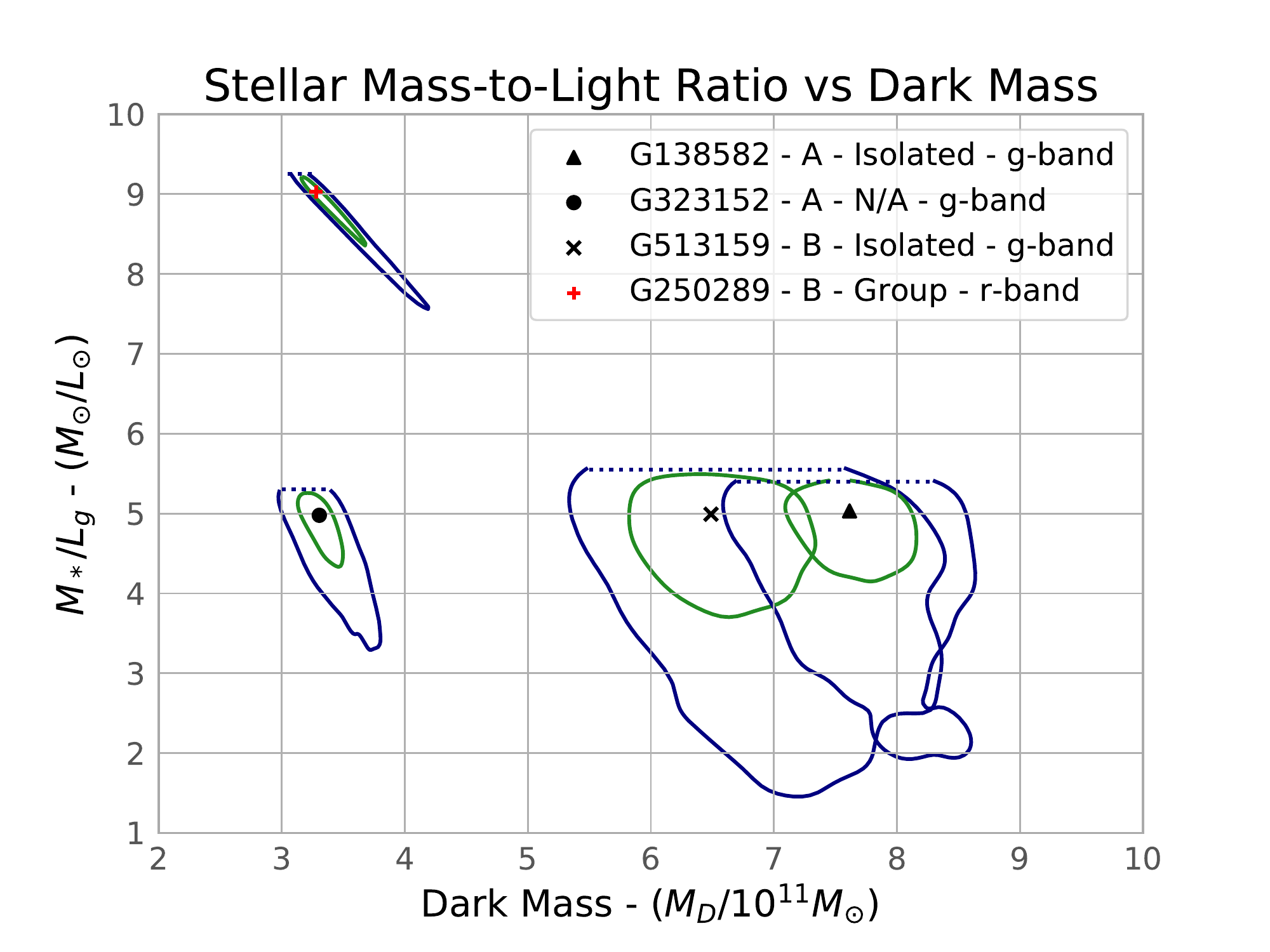}
    \caption{Stellar mass-to-light ratio ($M_*/L_g$) in g-band and dark mass integrated within the Einstein radius of each of the four best-fit models described in Section \ref{sect:model_results} and Table \ref{tab:final_four_results}. Dashed lines at the $2\sigma$ contours indicate boundaries corresponding to upper constraints placed on the mass-to-light ratio of the models (see Section \ref{sect:priors}). Legend and marker information are the same as in Figure \ref{fig:sm_dm}.}
    \label{fig:mls_fdm}
\end{figure}

We show the four highest-quality models for comparison. Figures \ref{fig:sm_dm}-\ref{fig:mls_fdm} show the four models in parameter space of interest to our study. All of the plotted quantities are taken within the Einstein radius (see Table \ref{tab:final_four_results}). Each model identified with a different marker, and B-grade group-member galaxy G250289 is indicated with a red marker to remind the reader that the final model fit utilized the r-band. Green and blue contours enclose $1\sigma$ (39\%) and $2\sigma$ (86\%) of the two-dimensional PDF respectively. With the small sample size shown here, we do not intend to address questions of assembly bias and galaxy formation mechanisms. These plots are intended to discuss the cleanest subset of our sample in the context of what can be considered more thoroughly in future work.

Figure \ref{fig:sm_dm} shows the integrated stellar mass and dark mass enclosed within the Einstein radius of the lens models. These are obtained by integrating over the S\'ersic stellar mass profile and elliptical NFW profile. The galaxies have total enclosed Einstein mass values of order $M_E \sim 3-8\times10^{11}$ $M_{\odot}$, which is is expected since lensing galaxies are typically quite massive. Assembly bias would show itself here as a trend where group central galaxies tend to have higher stellar mass than isolated galaxies at the same dark matter halo mass. The only group-member galaxy, G250289 (marked with a red cross), lies in one of the smaller dark matter halos and has the highest stellar mass. G323152 (marked with a black circle, not listed in GAMA {\sc GroupFinding} catalog) has a similar dark mass and about one-fifth of the stellar mass compared to G250289. Conversely, the two isolated galaxies have the highest dark masses and the lowest stellar masses. The higher stellar mass in G250289 could have more to do with effects from the difference in r-band and g-band S/N than a physical interpretation. With so few data, it is difficult to determine how much of an effect this difference has on the results of the models.

The total lensing (Einstein) mass enclosed within the Einstein radius is generally well-constrained. We want our models to further constrain the fraction of this lensing mass that is dark matter. The fraction of dark matter is not directly constrained by a model prior but is very sensitive to the assumed forms of mass and light profiles and the constraints placed upon those. Figure \ref{fig:lum_fdm} shows the g-band integrated luminosity and dark matter fraction (both calculated within the Einstein radius) for each model. The uncertainties in the fraction of dark matter along the x-axis are relatively small, which is surprising given the inherent degeneracy of stellar and dark mass in lens modeling. The small parameter space explored could indicate a lack of flexibility of the models' stellar and dark mass profile priors, perhaps an excessive constraint or weighting toward one mass component over the other. As discussed in Section \ref{sect:priors}, the careful selection of priors can be challenging. Compare again G250289 (red cross) and G323152 (black circle), which have similar enclosed dark masses. Even corrected (scaled down) from r-band to g-band, the enclosed luminosity for G250289 is 2-4 times greater than the other three, which could be why the model attributes a higher fraction of the Einstein mass to the stellar component. These models have relatively similar total Einstein masses enclosed within similar Einstein radii around 1.2-1.7 arcsec. G250289 has an Einstein radius of $\sim1.5$ arcsec that is typical and within the range of the other model values, so additional luminosity and stellar mass is not a result of an overextended radius of integration. 

Figure \ref{fig:mls_fdm} shows the g-band stellar mass-to-light ratio compared to the enclosed dark mass. Dotted lines at the 2$\sigma$ contours indicate upper constraints placed on the mass-to-light ratio. This figure completes our discussion of the degeneracy. In summary, the model has two options for attributing the lensing mass: (i) To favor the stellar component, a higher stellar mass can be the result of a heavier stellar population, and (ii) conversely, a very large, centrally concentrated dark matter halo can make up for a lower stellar mass and luminosity. This is all expected. Bounds of integration for luminosity, stellar mass, and dark mass are all dependent on the measure of the Einstein radius, which is in turn dependent on the total mass which the model is attempting to parse into stellar and dark components. It is a complex problem with degenerate variables that is only constrained by meaningful priors. In our cleanest subsample, the most identifiable differences occur for the candidate that was modeled from a different photometric bandpass. This is another experimental design decision based on limitations of the data that significantly affected our ability to analyze the resulting models. Thus, even our best models suffer.

\section{AUTOZ Considerations Post-Modeling}
\label{sect:autoz_quality}

\begin{figure}
    \centering
    \includegraphics[width=\columnwidth]{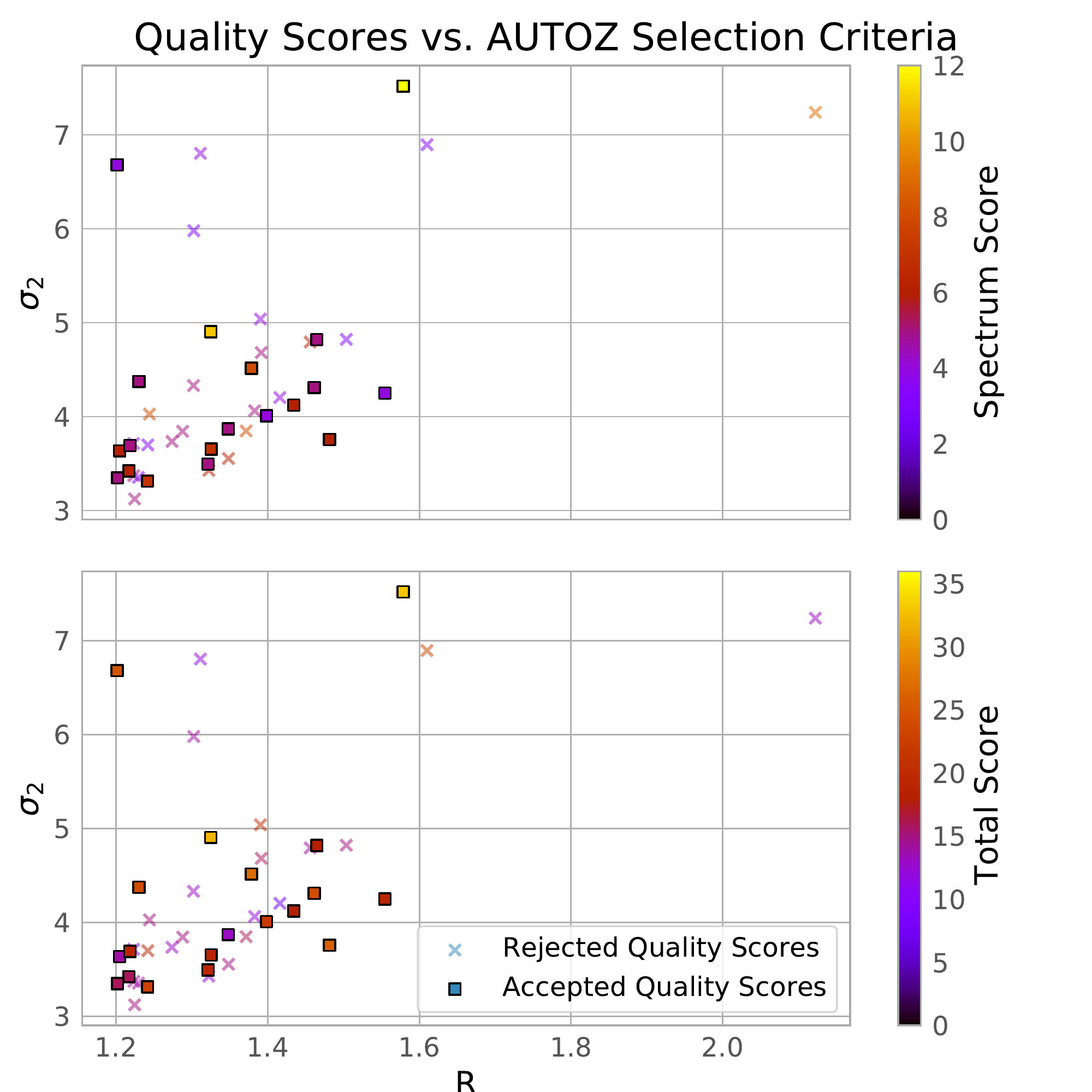}
    \caption{\textit{Upper}: Spectrum quality scores and \textit{Lower}: total score shown as scatter plot color variations scaled with the colorbar on the right of each plot. The axes of the scatter plot are the selection criteria from Figure \ref{fig:sigma_selection}. Vertical axes are the second-highest cross-correlation peak $\sigma_2$, and horizontal axes are the $R$ parameter. Dark purple colors indicate poor scores, with higher scores at orange and yellow. Little correlation can be seen between these parameters and the results of the models or their subjective spectrum scores. In the upper plot, four low-scoring spectra with high $\sigma_2$ correspond to overlapping emission lines.}
    \label{fig:quality_scores_vs_autoz}
\end{figure}

Here we summarize the results of our quality control procedure described in Section \ref{sect:quality_scoring} and give several recommendations for removing contaminants. Appendix \ref{sect:spec_qual_ctrl} gives a more detailed breakdown of the candidates and models in the context of the cases discussed in Section \ref{sec:autoz_cases}.

We discovered some failure modes for our method of utilizing the {\sc Autoz} algorithm for background source redshift determination. Revisions to our initial procedure removed about half of the candidates. Few single cases should be excluded without consideration; for example, some redshift determinations were kept even though there appeared to potentially be a falsely attributed line. However, absorption and emission lines must both be checked for overlap regardless of the template configuration type, and template type should be questioned with this information. Several of the lens ELG galaxies turned out to have prominent absorption features at the same redshift. The binary identification of PG and ELG, as we have done here, is an oversimplification that hinders both our classification and understanding of the spectra and expected galaxy properties. 

The two A-grade candidate models were acquired with {\sc Autoz} redshift matches that we consider to be cases that deserve extra care (where the foreground galaxy is best matched with an ELG template). This is interesting because one may be tempted to cut these cases entirely in order to obtain a clean and fairly homogeneous sample (i.e. large elliptical galaxy lensing a bright emission line galaxy, PG+ELG). We find the more careful consideration of other possible cases to be fruitful. If we had adopted a selection that focused only on configurations where the primary template matched a passive lensing galaxy, then $\sim20\%$ of our final selection, including the two highest-scoring models, would have never been considered. Care should be taken in order to reap the benefits of this expanded population while minimizing contamination.

In the big picture, the improvement of spectral quality in wide-field surveys is essential for making this work in an automated way over large sample sizes, but an automated redshift algorithm like {\sc Autoz} could be optimized for background source redshift determination. Our subjective quality scores show little correlation to the {\sc Autoz} output parameters that we used for the initial selection, as shown in Figure \ref{fig:quality_scores_vs_autoz}. The two axes show the {\sc Autoz} selection parameter space, composed of the second cross-correlation peak $\sigma_2$ and the $R$ parameter. Three of the candidates with the highest $\sigma_2$ show very poor spectrum scores because they are instances of overlapping emission lines. This suggests that a higher threshold for $\sigma_2$ may not actually yield higher-quality background source redshift matches. We now introduce other options for maintaining a clean sample selection with {\sc Autoz} while expanding the sample size in future works.

\subsection{Recommendations to Remove Contaminants from AUTOZ Selection}\label{section:recommendations}

\begin{figure}
    \centering
    \includegraphics[width=\columnwidth]{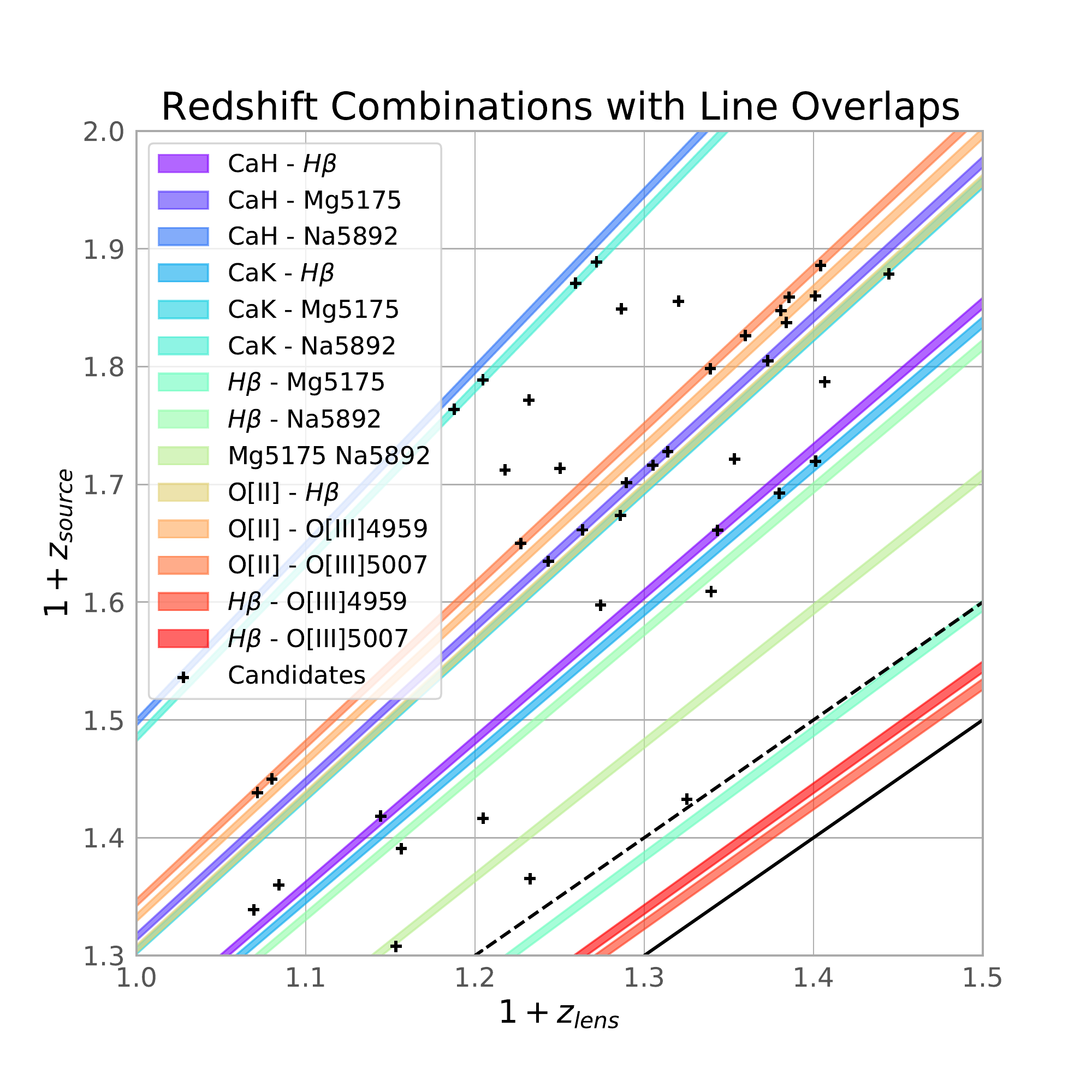}
    \caption{Combinations of foregrounds len and background source redshift that result in overlapping important emission or absorption line features. Colored line-regions are calculated with Equation \ref{equation:overlap} and indicate parameter space where the labeled line features will overlap. Each is labeled in the legend as "source feature - lens feature". Using these functions identifies 19 of the 20 overlaps in the 42 candidates and all 6 that made the final selection of 19. The region in the lower right between the solid and dashed black lines shows the selection criterion utilized in the initial {\sc Autoz} selection, which removed candidates where the source redshift was within 0.1 of the lens redshift.}
    \label{fig:overlapping_functions}
\end{figure}

\begin{figure}
    \centering
    \includegraphics[width=\columnwidth]{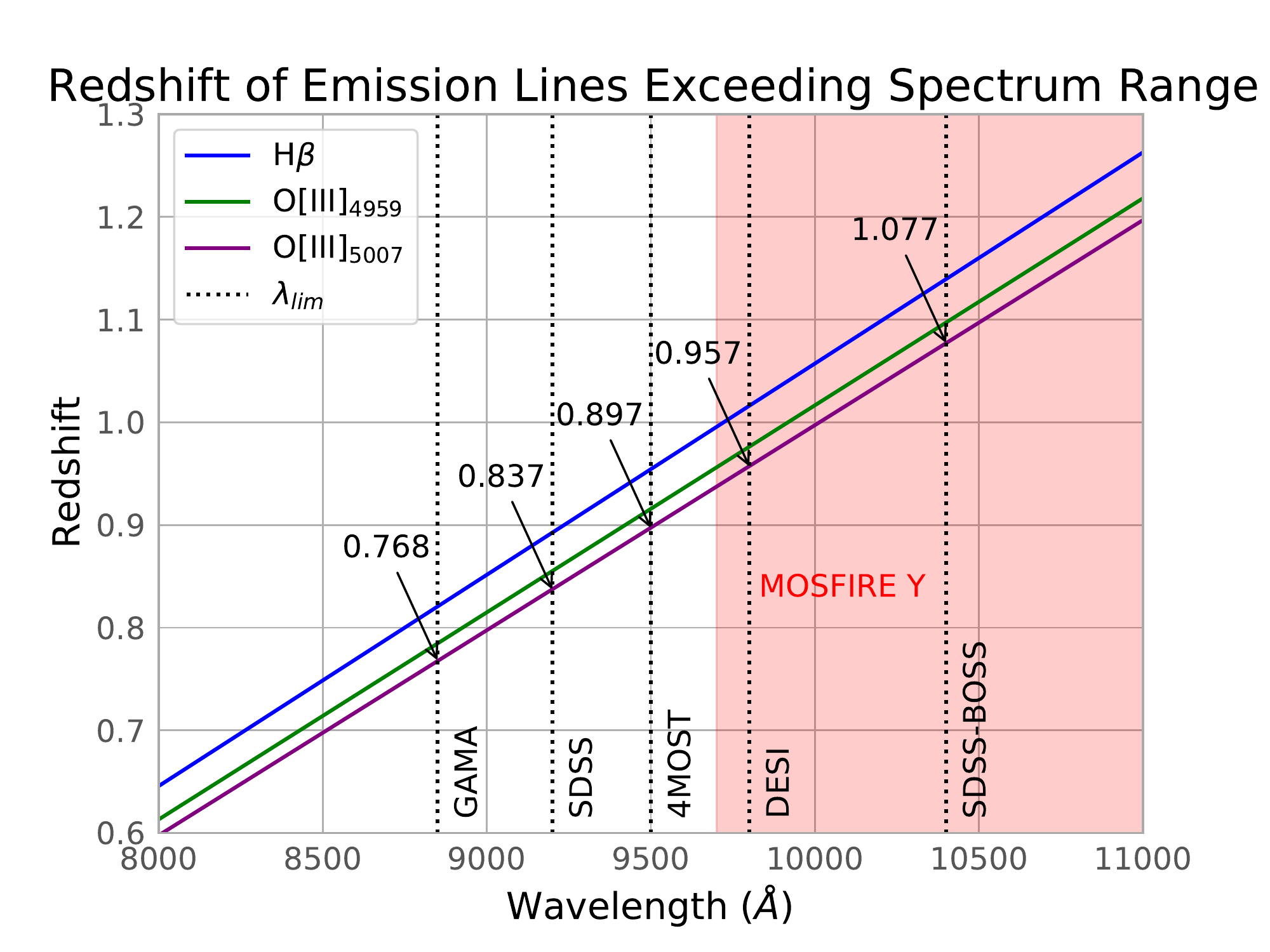}
    \caption{Blue, green, and purple lines show the redshift of restframe H$\beta$ and [O III]$\lambda\lambda$4959,5007. Dotted vertical lines show the upper wavelength limit of various spectroscopic surveys, and the overlaid arrows point to the redshift at which the first of these three lines will disappear in the survey observations. The red shaded region indicates the very near-IR coverage of MOSFIRE Y-band (0.97-1.12 $\mu$m) that could also potentially reveal background source emission lines at around z$\sim$1 and greater.}
    \label{fig:wavelengths_exceed}
\end{figure}

One could quite easily remove contaminants during the automated selection. Each of these recommendations pays particular attention to the redshifts of emission line galaxies in both the foreground lens and background source positions. The two most convincing (A-grade) spectra and models were cases of ELG+, so we want to remove as many contaminants as possible without the blanket removal of either of these cases. In order to maintain the applicability of this procedure to an even larger set of data than is considered here, we recommend the following selection criteria be implemented when adopting automated redshift determinations:

\begin{enumerate}
    \item \textbf{Remove ELG+ELG and PG+PG matches where emission or absorption lines redshift to overlapping wavelengths between foreground lens and background source.} One can calculate lens and source redshift combinations that result in overlapping observed emission lines similarly to the procedure described in \cite{Holwerda15}. The following equation defines a region of parameter space between a lower and upper linear function of $(1+z_{\mathrm{source}})$ to $(1+z_{\mathrm{lens}})$. Within this region, an overlap will occur for a given pair of restframe emission line wavelengths:
    
    \begin{equation}\label{equation:overlap}
        \left| \frac{1+z_{\mathrm{source}}}{1+z_{\mathrm{lens}}} - \frac{\lambda_{r,\mathrm{lens}}}{\lambda_{r,\mathrm{source}}} \right| < \frac{A}{R}
    \end{equation}
    
    \noindent where $z_{\mathrm{source}}$ and $z_{\mathrm{lens}}$ are the source and lens redshifts, $\lambda_{r,\mathrm{lens}}$ and $\lambda_{r,\mathrm{source}}$ are the restframe wavelengths of emission lines from lens and source, $R$ is the spectral resolution of the instrument, and \textit{A} is a coefficient that widens the range of exclusion for potential overlapping features. The equation implicitly accounts for the dependence of resolution on observed wavelength. Figure \ref{fig:overlapping_functions} shows the regions where a redshift combination of foreground lens and background source results in overlapping lines given GAMA's spectral resolution of $R\sim1300$ and $A=4$ (i.e. overlapping emission lines are closer than 4 times the smallest resolvable wavelength difference). This prescription identifies most of the cases of overlap that were flagged by direct visual inspection of the spectra. We retained one of them with the lowest possible D-grade because its SDSS-BOSS spectrum showed fairly reasonable source H$\beta$ and [O III] couplet lines at higher-wavelength that were not in the range of the GAMA spectrum.
    
    \item \textbf{Remove +ELG configurations where H$\beta$ and [O III] couplet emission lines are redshifted beyond the wavelength range of the observation.} Table \ref{table:survey_redshift_limits} and Figure \ref{fig:wavelengths_exceed} show the redshift limits beyond which H$\beta$ and [O III]$\lambda\lambda$4959,5007 lines will be above the survey upper wavelength limit for several optical spectroscopic surveys.  Spectroscopy in the 1-$\mu$m range is necessary in order to detect the emission lines from sources at z$\sim$1 and will become more important for modeling lenses with foreground lens redshifts higher than z$\sim$0.5. DESI and the upcoming 4MOST have higher resolution and cover extended optical wavelength ranges that correspond to these redshifts, which will reduce the significance of this problem. Background source redshifts could also be assessed by looking for emission lines in very near-infrared (e.g. MOSFIRE Y-band, 0.97-1.12 $\mu$m) spectra, where possible. In principle, template-based automated redshift identification could be run for each separately, which would negate some of the difficulties inherent to identifying the two distinct signals within the single observation. The obvious negative to this option is the need for additional observations.
    
	\item \textbf{Remove ELG+ matches with any of the following characteristics: (a) low foreground lens redshift (less than z$\sim$0.2 for our sample), (b) ELG+PG configuration, and (c) primary redshift match to the background source.} ELG+ configurations with low foreground lens redshifts suffered from several failure conditions in addition to being a less-likely configuration than PG+. For GAMA, the noisy short-wavelength end of the observed spectral range corresponds to emission lines with $z_{\mathrm{lens}}$ less than $\sim$0.2, leading to mistaken classification of noise peaks as emission lines. While this is specific in part to GAMA's wavelength-specific spectral performance, several of these low-redshift foreground lens matches are also ELG+PG configurations, of which almost all were removed in quality scoring. The one remaining candidate had the lowest score of all those accepted. Perhaps even more condemning is the fact that many of these low-redshift ELG+ foreground lens matches were also cases where the background source was assigned the primary redshift. This trio of doubtful cases coincided for several of the candidates that were removed from our sample during quality scoring.  
	
\end{enumerate}

\begin{table*}
\begin{center}
\begin{tabular}{l l l l l}
\hline
Survey & $\lambda_{lim}$ ({\AA}) & $z_{\mathrm{source}}$& & \\
& & H$\beta$ & [O III]$\lambda4959$ & [O III]$\lambda5007$ \\
\hline
AAT (GAMA/DEVILS) & 8850 & 0.821 &  0.785 & 0.768 \\
SDSS (original) & 9200 & 0.893 & 0.855 & 0.837 \\
4MOST (lo-res) & 9500 & 0.954 & 0.916 & 0.897 \\
DESI & 9800 & 1.016 & 0.976 & 0.957 \\
SDSS-BOSS & 10400 & 1.139 & 1.097 & 1.077 \\
\hline
\end{tabular}
\end{center}
\caption{Five spectroscopic surveys and their upper wavelength limits limits. The right three columns show the source redshift at which the given emission line will be redshifted beyond the upper wavelength limit of the observation.}
\label{table:survey_redshift_limits}
\end{table*}%

\begin{table*}
\begin{center}
\begin{tabular}{l l l l l l l}
\hline
 GAMA ID & 	$z_{\mathrm{lens}}$ 	& $z_{\mathrm{source}}$ & $\sigma_{\mathrm{lens}}$ & $\sigma_{\mathrm{source}}$ & Type & Grade \\

\hline
323152 & 0.353 & 0.722 & \textbf{7.52} & \textbf{11.32}  & PG+ELG &	A \\
262836 	& 0.418 & 0.144 & \textbf{3.87} & \textbf{10.23} & ELG+PG &	D \\

\hline
\end{tabular}
\end{center}
\caption{Two models with {\sc Autoz} primary redshift template match to the background source and secondary match to the foreground lens ($\sigma_{\mathrm{lens}}<\sigma_{\mathrm{source}}$). Type refers to the foreground+background configuration of galaxy templates. Grade is an evaluation of the quality of the fit to the image according to the scheme outlined in Table \ref{table:quality_grades}. G323152 is one of the highest scoring models in this study.}
\label{table:primary_background}
\end{table*}%

\section{Discussion}\label{sect:discussion}

\subsection{GAMA Environment}
\label{sect:environment}

The sample of 42 KiDS lens candidates and the subsample of 19 with accepted grade A-D models are both close to evenly split between group-member and isolated galaxies according to the GAMA {\sc GroupFinding} metrics. 22 (8 graded) are associated with groups and 19 (9 graded) are isolated. (G323152) is not represented in GAMA group catalogs.

We note that most strong lensing galaxies should be the most massive galaxies in their halo, either in a group or in isolation. Figure \ref{fig:group_distance} shows the rank of the proximity of the object to the center of mass of the group relative to other group members, with 1 indicating the closest or center-most galaxy. Most of the candidates shown here are group central galaxies, but our models failed for 11 of those. On the other hand, 4 of the 5 candidates that are not the central galaxy of their group were accepted and given grades, comprising 40\% of the graded group-galaxy members. Low scores for group member galaxies could mean that their identification as lenses is a false-positive given their proximity to other significantly large galaxies. Alternatively, if these are lenses, the modeled mass structure of the lensing galaxy as a single light-mass component and assumption of its presence in the center of the dark matter halo may not be accurate enough to reproduce the lensing observables. If this were the case, one might expect the group centrals to be more easily modeled than the subdominant or "satellite" galaxies with ranks of 2 or greater, which is not apparent in Figure \ref{fig:group_distance}. The two B-grade group galaxies are ranked 1 and 2, indicating that one of them is a central while the other has at least one companion that is competing for dominance in the group. The rank 2 group member is G62734, which was removed from the final selection because its dark matter content was poorly constrained. This could be a result of this galaxy's distance from the center of the group mass.

\begin{figure}
    \centering
    \includegraphics[width=\columnwidth]{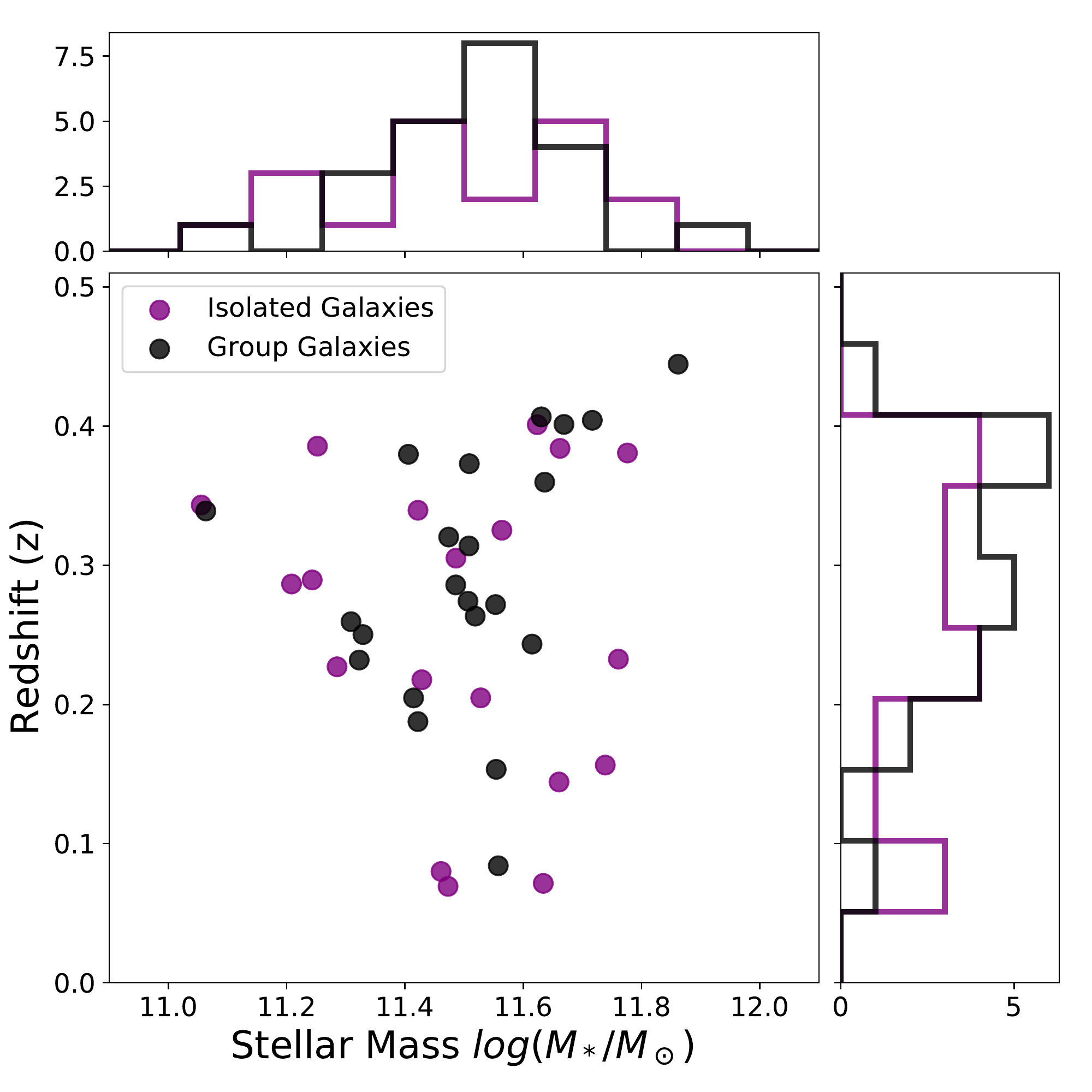}
    \caption{Redshifts and stellar masses of {\sc Autoz} sample from GAMA-DR3 {\sc StellarMassesLambdar v20} \citep{Taylor16} determined by stellar population and separated by group-member and isolated galaxies according to GAMA team internal {\sc GroupFinding} catalogs \citep{Robotham11}. There is no clear distinction between the subsamples in either observable.}
    \label{fig:group_mass_redshift}
\end{figure}

\begin{figure}
    \centering
    \includegraphics[width=\columnwidth]{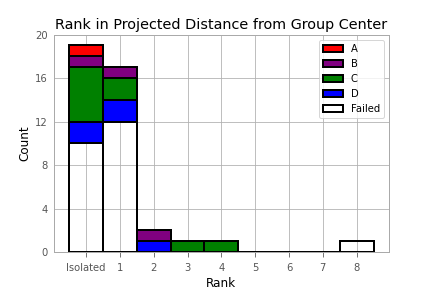}
    \caption{Stacked histogram. Quality scores for 41 of 42 candidates in reference to data from GAMA {\sc GroupFinding} catalogs (one candidate does not have environment data). Location on the x-axis distinguishes "Isolated" from group member galaxies, which are further separated by the rank in projected distance from the center of mass of the group. A rank of 1 indicates that the lens candidate is the central galaxy of the associated group. Colors indicate the quality grade of models, with no color indicating models that were not accepted.}
    \label{fig:group_ranks}\label{fig:group_distance}
\end{figure}

Compared with the SLACS study in \cite{Treu09}, in which 12 of 70 (17\%) were associated with groups, our KiDS/GAMA strong lens candidate sample and selected subsample of 19 models is more highly represented by group-member galaxies. Definitions of group membership based on environmental parameters are not the same between these studies. The nearly 50/50 split between group-member and isolated galaxies in our sample does not necessarily support or dispute a preference for overdense environments by lensing (and all massive) elliptical galaxies. However, the high completeness of GAMA compared with SDSS may instead suggest that our sample minimizes the apparent environmental preference. The distinction of group association here could be affected by selection bias, as those designated as isolated could in fact be groups with satellite members beyond the GAMA flux limit. If this were the case, there should be a systematic bias in isolated galaxies toward higher redshift. Figure \ref{fig:group_mass_redshift} shows that neither subsample of group member or isolated candidates is significantly distinguished in redshift or stellar mass. 

With more data and better measurements than we have accomplished here, one may be able to compare observations to the scatter in the upper plot of Figure 9 of \cite{Zehavi18}, where for fixed dark halo mass, higher stellar-mass galaxies tend to exist in denser environments. Note that the modeled mass components here are calculated within the Einstein radius and not the full extent of the galaxy. The majority of the dark halo component should extend well beyond the stellar halo, and these high-mass lensing galaxies are more likely to exist in more massive dark matter haloes  ($\mathrm{log}(M_h/h^{-1}M_{\odot})\sim12-14$) where the suggested environmental trend is less supported. The precision and numbers required to test assembly bias will require more refinement of the methods discussed in this study as well as the power of more sophisticated surveys to come.

\subsection{Future Work: a Place for Ground-Based Observations}\label{section:space_imaging}

\begin{figure*}
    \centering
    \includegraphics[width=0.45\textwidth]{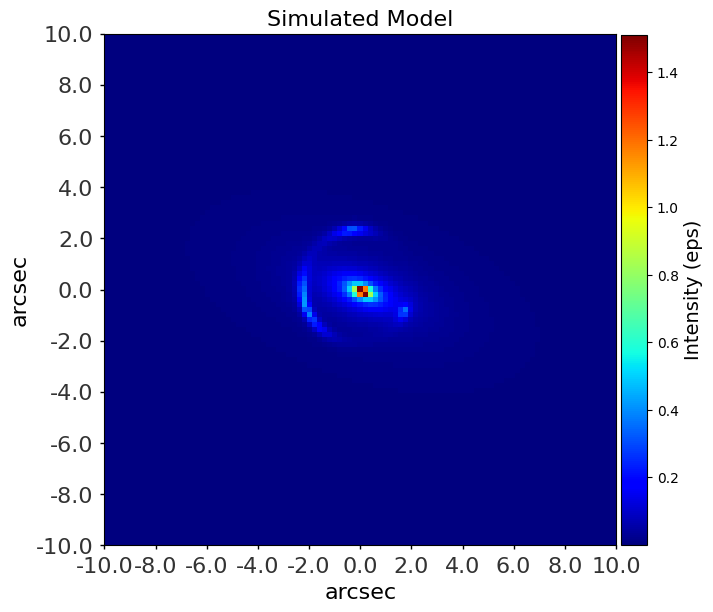}
    \includegraphics[width=0.45\textwidth]{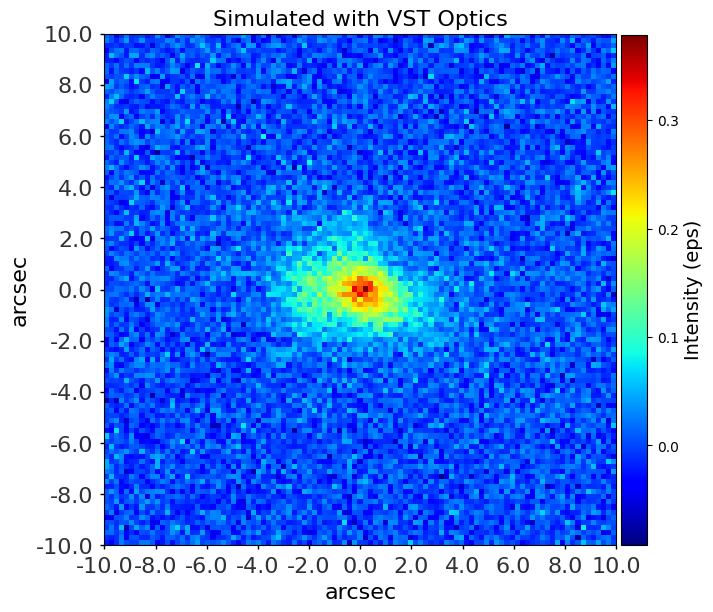}
    \includegraphics[width=0.45\textwidth]{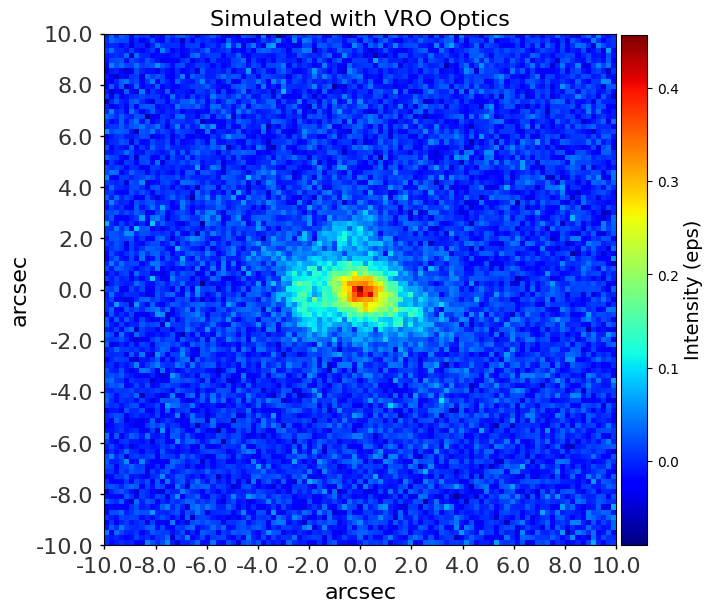}
    \includegraphics[width=0.45\textwidth]{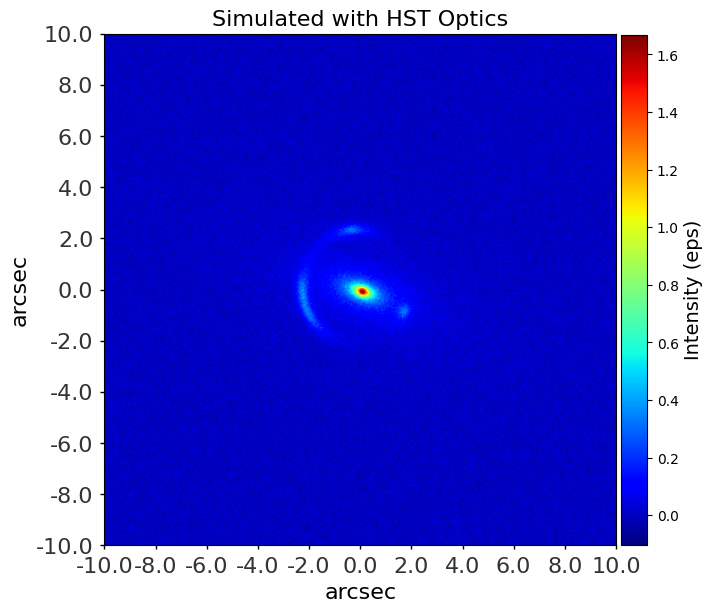}
    \caption{\textit{Upper left}: Maximum log-likelihood model for G3629152 shown with pixel-scale 0.2 arcsec/pixel. Other images are simulated with identical background sky and convolved with optics of \textit{upper right}: VST (r-band PSF 0.65 arcsec, pixel scale 0.2), \textit{lower left}: LSST at Rubin (PSF 0.5, pixel scale 0.2), and \textit{lower right}: ACS on HST (PSF 0.1, pixel scale 0.05). Imaging from space-based observatories or AO would allow for better model-fitting and tighter uncertainties for future efforts.}
    \label{fig:simulated_lenses}
\end{figure*}

Realistic lens modeling by fitting mass and light profile parameters is a complex problem with a large number of parameters. With even the highest-quality ground-based imaging offered by the likes of the Kilo-Degree Survey, the angular resolution is insufficient to constrain the \textit{individual} model solutions to levels where one can make strong inferences about the \textit{individual} lens galaxies. There are simply too many solutions that fit the image to a high probability, which inflates the uncertainty to levels that make it difficult for one to draw conclusions from the inferred quantities. These uncertainties on a single lens can be significantly constrained with the level of imaging afforded by AO or space-based instruments. Figure \ref{fig:simulated_lenses} shows a model solution for one of the lens models after being simulated with the optics for three observatories: (i) VLT Survey Telescope (VST) used for KiDS, which was the instrument that collected the original image, (ii) LSST at the Vera Rubin Observatory (VRO) representing the next generation of ground-based observatories, and (iii) Advanced Camera for Surveys (ACS) on the Hubble Space Telescope. The same model-fitting procedure applied to HST images or observations with adaptive optics (AO) of the same lensing galaxies would result in error estimates an order of magnitude better than the results we achieve here. Alternatively, future systematic modeling of orders of magnitude more ground-based, lower-resolution observations (as we expect to achieve with observatories like the VRO) can result in similar precision. Constraints at the population level (made possible with these larger sample sizes) can enhance higher-resolution individual measurements through Bayesian hierarchical frameworks. Our work demonstrates the value of wide-field, lower-resolution surveys as a complementary tool to the expensive and hyper-competitive observing campaigns that are the default for strong lens studies.

The next generation of spectroscopic surveys is already underway, e.g. the DEVILS deep survey on the AAT \citep[]{Holwerda21, Davies18} and the DESI redshift survey. One can expect increased numbers of spectroscopic lensing candidates as well as opportunities to identify the redshift of the potential background source. 
More comprehensive spectroscopic surveys are being planned with the 4MOST instrument \citep{de-Jong12a,Depagne14}. These planned surveys include extra-galactic ones such as the two-tiered Wide Area Vista Extragalactic Survey \citep[WAVES,][]{Driver19}, the Optical, Radio Continuum and HI Deep Spectroscopic Survey (ORCHIDSS, Duncan et al. \textit{in prep.}), and a cosmological low-S/N wide-area survey \citep[CRS,][]{Richard19}. These 4MOST surveys are expected to achieve high completeness in their target fields and yield a boon of spectroscopically confirmed strong lensing systems with the same advantages exploited by the procedure outlined here at better spectral resolution and wider fields.

Identifications of strong gravitational lenses through imaging are also expected to increase in the near future with observations by the likes of the Vera Rubin Observatory, Euclid, and the Roman Space Telescope, in addition to improved machine learning techniques. 
Following the discussion in \cite{Knabel20}, one expects the selection functions of the spectroscopic surveys and these optical and near-infrared imaging surveys to show a limited improvement in overlap. The analysis presented here, however, shows that in the overlap a useful subset of strong lenses can be utilized for modeling by imaging and spectroscopy combined. 

The lenses discussed here, and in future similar ground-based efforts, are also ideal candidates for deeper follow-up observation with higher-resolution imaging and spectroscopy. These follow-up observations could include Integral Field Unit (IFU) observations to measure the stellar population characteristics across the elliptical, 
chart the foreground lens galaxy kinematics, as well as study the background source light and stellar population characteristics. These considerations are more aligned with and have been sufficiently described in the existing literature and will not be discussed further here.

\section{Conclusions}
\label{s:conclusions}

We arrive at the following conclusions from our analysis of strong lens candidates in the Kilo-Degree Survey using {\sc Autoz} and {\sc PyAutoLens}:
\begin{enumerate}
    \item Meaningful strong-lens studies can be conducted by applying lens-modeling methods such as those we have outlined here to large imaging and spectroscopic surveys.
    \item Automated template-matching redshift algorithms like {\sc Autoz} can be utilized to determine reliable background source redshifts required for lens modeling. Careful consideration should be taken in cleaning the algorithm's selection, following the recommendations outlined in Section \ref{section:recommendations}.
    \item Limits of optical resolution in large ground-based surveys present significant challenges to the uniqueness of solutions in our Bayesian modeling of individual strong lenses.
    \item As sample sizes grow, refinements to these techniques can produce lensing measurements in \textit{quantities} that will offer considerable statistical power. This approach is complementary to the more detailed modeling of individual lenses that is possible with deeper and higher resolution observations.
\end{enumerate}

\section{Acknowledgements}

AHW is supported by an ERC Consolidator Grant (No. 770935). SK acknowledges NASA Kentucky, National Science Foundation (NSF), University of Louisville, and University of California, Los Angeles, for financial and technical support. The material of this work is based upon work supported by NASA Kentucky under NASA award No: 80NSSC20M0047. This material is based upon work supported by the National Science Foundation Graduate Research
Fellowship Program under Grant No. 2021325146. Any opinions, findings, and conclusions
or recommendations expressed in this material are those of the author(s) and do not necessarily reflect
the views of the National Science Foundation.

\section*{Data Availability}

KiDS images and data used in this paper are available from the Astro-WISE Database Viewer Web Service (\url{dbview.astro-wise.org}). LinKS-specific data can be found at the LinKS website (\url{https://www.astro.rug.nl/lensesinkids/}). The GAMA {\sc Autoz} catalog is available from GAMA-DR3 website (\url{http://www.gama-survey.org/dr3/}, {\sc AATSpecAutozAll}, {\sc SpecAll},  {\sc LamdarStellarMasses},  {\sc SersicCatSDSS}, and  {\sc kcorr\textunderscore auto\textunderscore z00} catalogs) and the team internal {\sc GroupFinding} catalog for the full GAMA fields will be made available in GAMA-DR4 (Driver et al. \textit{in preparation}).

\section*{Software Citations}

This work uses the following software packages:

\begin{itemize}

\item
\href{https://github.com/astropy/astropy}{\texttt{Astropy}}
\citep{astropy1, astropy2}

\item
\href{https://bitbucket.org/bdiemer/colossus/src/master/}{\texttt{Colossus}}
\citep{colossus}

\item
\href{https://github.com/dfm/corner.py}{\texttt{corner.py}}
\citep{corner}

\item
\href{https://github.com/joshspeagle/dynesty}{\texttt{dynesty}}
\citep{dynesty}

\item
\href{https://github.com/matplotlib/matplotlib}{\texttt{matplotlib}}
\citep{matplotlib}

\item
\href{numba` https://github.com/numba/numba}{\texttt{numba}}
\citep{numba}

\item
\href{https://github.com/numpy/numpy}{\texttt{NumPy}}
\citep{numpy}

\item
\href{https://github.com/rhayes777/PyAutoFit}{\texttt{PyAutoFit}}
\citep{pyautofit}

\item
\href{https://github.com/Jammy2211/PyAutoLens}{\texttt{PyAutoLens}}
\citep{Nightingale2015, Nightingale2018, pyautolens}

\item
\href{https://www.python.org/}{\texttt{Python}}
\citep{python}

\item
\href{https://github.com/scikit-image/scikit-image}{\texttt{scikit-image}}
\citep{scikit-image}

\item
\href{https://github.com/scikit-learn/scikit-learn}{\texttt{scikit-learn}}
\citep{scikit-learn}

\item
\href{https://github.com/scipy/scipy}{\texttt{Scipy}}
\citep{scipy}

\end{itemize}

\bibliographystyle{mnras}
\bibliography{shawn_bibliography, shawn, james} 


\appendix

\section{Preparing Data for Modeling}
\label{s:prepdata}

Images and weight maps are $101\times101$ pixel ($\sim20\times20$ $arcsec^2$) cutouts from coadded images of KiDS tile observations acquired from the publicly available Astro-WISE Database Viewer Web Service\footnote{\url{dbview.astro-wise.org}}. g- and r-band images are cut out centered on the object's RA and DEC, recentered to the brightest pixel in the central (lens) galaxy light profile, and converted to eps (electrons per second) for modeling. KiDS image pixel values in the Astro-WISE Database are given in calibrated flux units relative to the flux corresponding to magnitude 0 and are converted to "brightness" units of electron counts by multiplying by the tile's average gain, which includes additional  factors necessary for this conversion. {\sc PyAutoLens} is by default set to be optimally utilized with units of electrons per second (eps), which is acquired by dividing by the exposure time (1800 seconds for r-band, 900 for g-band). 

{\sc PyAutoLens} requires input of the PSF and noise map for each image. The inverse square root of the weight map corresponding to the cutout image gives the rms noise, which is converted to electron counts and squared to recreate the background sky. We then add this image to the corresponding cutout image and take the square root to give the noise map, after which we convert to eps. We generate a Gaussian PSF for each image from the average FWHM PSF for each image. 

We next mark pixel-positions of the distorted images of the lensed background source in each image, when visible, using a GUI distributed with {\sc PyAutoLens}. During a lens model fit, {\sc PyAutoLens} casts aside all mass models where these image-pixel positions do not trace within a designated threshold of one another in the source plane. This narrows the parameter space that is searched and ensures that the model fits the observed image features of interest.

We generate three masks for the three searches with each candidate: (i) \textit{lens mask} --- a circular aperture tailored to show only the lens galaxy (on the order of but usually slightly less the effective radius, typically around 1-1.3 arcseconds); (ii) \textit{source mask} --- a circular annular aperture showing only the light we determine to be the lensed background source features (with inner radius about the size of the circular lens mask and outer radius around 3 arcseconds); and (iii) \textit{full mask} --- a circular aperture of typically around 3 arcseconds that includes most of the light from the lens and source features and masks as many peripheral contaminants as possible.

\section{Lens Modeling Pipeline}\label{sect:pipeline}

This section details continues the description of our lens modeling methods summarized in Section \ref{sect:pyautolens_description}. Through experimentation, we have designed a pipeline composed of a chain of three {\sc Dynesty} searches that we use as a template for fitting each lens. The variety of lensing configurations, image quality, etc. force us to tailor aspects of each model-fit individually, in particular alternating the masks that segment the foreground lens light and distorted background source light. In order to institute the least bias possible, we allow the models to probe a wide range of possible solutions for each parameter. The shape of the prior distribution has significant effects on the performance of the search. We use uniform, log uniform, and Gaussian functions depending on the parameter and informative auxiliary observations. In the following sections, we describe this three-step automated pipeline, where from here on we refer to a "search" as a model-fit performed by the non-linear search {\sc Dynesty}. Each subsequent search in the chain has more complexity in the form of additional parameters, which we balance in computational time by passing priors from previous search outputs. For each non-leanear search in the chain, the priors are described in Table \ref{table:phases}, and the {\sc Dynesty} settings are given in Table \ref{table:phases_settings}.

\begin{table*}
\begin{center}
\begin{tabular}{l l l l l l}
\hline

& $\#$ Free & & & & \\
Search & Parameters & Fit & Profile & Prior & Probability Density Function \\

\hline
1 & 7 & Lens Light & Elliptical S\'ersic & Center (y, x) & Uniform (-0.3 - 0.3 arcsec) \\
& & & & Elliptical Comps ($\epsilon_1$, $\epsilon_2$) & Gaussian (mean = 0.0, $\sigma$ = 0.3) \\
& & & & Intensity & Log Uniform ($10^{-6}$ - $10^{6}$ eps) \\
& & & & Effective Radius & Gaussian (GAMA-DR3 r mean and $\sigma$ \\
& & & & & \hspace{0.2cm}arcsec \textit{or} SLACS 7 kpc $\pm$ 3.3 at lens \\
& & & & & \hspace{0.2cm}distance, upper limit = mean + $3\sigma$ ) \\
& & & & S\'ersic Index & Uniform (0.5 - 8.0) \\

\hline
2 & 10 & Lens Light & Elliptical S\'ersic & Center (y, x) & Prior Passed from Search 1 (fixed) \\
& & & & Elliptical Comps ($\epsilon_1$, $\epsilon_2$) & Prior Passed from Search 1 (Gaussian) \\
& & & & Intensity & Prior Passed from Search 1 (Gaussian) \\
& & & & Effective Radius & Prior Passed from Search 1 (Gaussian) \\
& & & & S\'ersic Index & Prior Passed from Search 1 (Gaussian) \\
& & Lens Stellar Mass & Elliptical Isothermal &  Center (y, x) & Paired to Lens Light Prior (fixed) \\
& & & & Elliptical Comps ($\epsilon_1$, $\epsilon_2$) & Paired to Lens Light Prior (Gaussian) \\
& & & & Einstein Radius & Gaussian (mean = 1.0, $\sigma$ = 0.5, \\
& & & & & \hspace{0.2cm} limit = 0 - 2.5 arcsec) \\
& & Source Light & Spherical Exponential & Center (y, x) & Uniform (-2.0 - 2.0 arcsec) \\
& & & & Intensity & Log Uniform ($10^{-6}$ - $10^{6}$ eps) \\
& & & & Effective Radius & Gaussian (7.5 kpc $\pm2.5$ at source \\
& & & & & \hspace{0.2cm} distance, upper limit = mean + $3\sigma$) \\

\hline
3 & 14 & Lens Stellar Light & Elliptical S\'ersic & Center (y, x) & Prior Passed from Search 2 (fixed) \\
& & and Mass & & Elliptical Comps ($\epsilon_1$, $\epsilon_2$) & Prior Passed from Search 1 (Gaussian) \\
& & & & Intensity & Prior Passed from Search 2 (Gaussian) \\
& & & & Effective Radius & Prior Passed from Search 2 (Gaussian) \\
& & & & S\'ersic Index & Prior Passed from Search 2 (Gaussian) \\
& & & & Mass-to-Light Ratio & Log Uniform (Limits calculated) \\
& & Lens Dark Mass & Elliptical NFW & Center (y, x) & Paired to Stellar Mass prior (fixed) \\
& & & & Elliptical Comps ($\epsilon_1$, $\epsilon_2$) & Gaussian (mean = 0.0, $\sigma$ = 0.3) \\
& & & & $\kappa_s$ & Uniform (0.0 - 1.0) \\
& & & & Scale Radius & Gaussian (calculated from SLACS-IV \\
& & & & & $\hspace{0.2cm}$mean and $\sigma$) \\
& & Source Light & Spherical Exponential & Center (y, x) & Prior Passed from Search 2 (Gaussian) \\
& & & & Intensity & Prior Passed from Search 2 (Gaussian) \\
& & & & Effective Radius & Prior Passed from Search 2 (Gaussian) \\
 
\hline
\end{tabular}
\end{center}
\caption{Details about model searches and priors for three-step lens model-fitting with {\sc PyAutoLens}. Each phase fits a number of free parameters that model light and mass profiles of the lens and source galaxies by exploring the parameter space according to the prior's probability density function. Parameters fit in Searches 1 and 2 are input as Gaussian or fixed priors for subsequent searches. "Elliptical Components" are related to the axis ratio and position angle as in Equations \ref{equation:epsilon1} and \ref{equation:epsilon2}. See Sections \ref{sect:pipeline} and \ref{sect:priors} for more details about searches, profiles, and priors.}
\label{table:phases}
\end{table*}%

\begin{table*}
\begin{center}
\begin{tabular}{l l l l l l l}
\hline
Search & \textit{n} live points & Evidence Tolerance & Steps per Walk & Acceptance Fraction & Positions Threshold & Sub-Grid Size \\

\hline
1 & 200 & 0.5 & 10 & 0.3 & N/A & 2$\times$2 sub-pixels \\
2 & 300 & 0.25 & 10 & 0.3 & 1.5 arcsec & 2$\times$2 sub-pixels \\
3 & 500 & 0.25 & 10 & 0.3 & 1.5 arcsec & 2$\times$2 sub-pixels \\
\hline
\end{tabular}
\end{center}
\caption{{\sc Dynesty} non-linear search settings for each of the three searches of model-fitting. These settings balance computational cost with a thorough exploration of parameter space. Relaxed settings (e.g. low \textit{n} live points and high evidence tolerance) are useful for expediting initial fits that inform later fits. The trade-off is less well-defined uncertainty and a chance that the global maximum likelihood fit has been missed in favor of a local one. See Section \ref{sect:pyautolens_description} for a thorough description of {\sc PyAutoLens} and {\sc Dynesty} search settings.}
\label{table:phases_settings}
\end{table*}%

\subsection{Search 1 --- Lens Light}

Search 1 is the simplest and quickest of the three searches and focuses on returning an accurate lens light profile. The subtraction of this modeled light from the observed image should then show the lensed features of the background source. This search fits an elliptical S\'ersic profile \citep{Sersic68},

\begin{equation}\label{equation:sersic_profile}
    \centering
    I(R) = I_e e^{-b_n[(\frac{R}{R_e})^{1/n} - 1]}
\end{equation}

\noindent where $R$ is angular radius from the center of the profile, $I_e$ is the intensity at the effective radius $R_e$, $b_n \approx 2n - 0.327$, and $n$ is the S\'ersic index.
{\sc PyAutoLens} generates an image from these parameters in the image-plane and fits to the observed r-band image. The purpose of this search is to infer a high likelihood S\'ersic lens light model, which serves two purposes for Search 2: (i) it provides a lens-light subtracted image and; (ii) it provides lens light priors that are passed to subsequent searches. Because the image is centered during pre-processing, the distribution can be initialized with fairly tight constraints. Elliptical components,  $\epsilon_1$ and $\epsilon_2$, are defined as 

\begin{gather}
    \epsilon_1 = \epsilon_y = \frac{1 + b/a}{1 - b/a} \sin{2\alpha} \label{equation:epsilon1}\\
    \epsilon_2 = \epsilon_x = \frac{1 + b/a}{1 - b/a} \cos{2\alpha} \label{equation:epsilon2}
\end{gather}

{\noindent}where \textit{b} and \textit{a} are the semi-major and -minor axes of the ellipse, and $\alpha$ is the position angle. The intensity is parametrized according to electrons per second and therefore takes a wide log-uniform distribution. The S\'ersic index prior covers a wide range of reasonable values with a uniform distribution.

For the r-band images, many of the lensed background source's features are positioned within the lens galaxy's effective radius. Search 1 therefore struggles to deblend the lens and source light, and the distorted arcs of background source light are attributed to the foreground lens galaxy. In some cases, this leads to a model solution that describes a lens light profile that is very large and very elliptical. To mitigate this systematic effect, we use the aforementioned \textit{lens} mask for Search 1 and constraints on the effective radius to assist the search to focus on fitting the lens light and not the source light. The residuals and uncertainties of this search therefore tend to be quite high. In fact, the residuals often outline the lensed images of the source itself, and the resulting maximum log-likelihood is lower than the value inferred in the second and third searches.

\subsection{Search 2 --- Lens Mass and Source Light}

Search 2 focuses on the light from the background source. This component is modeled as a spherical exponential light profile in the source plane defined at the source redshift. The exponential light profile corresponds to the simple $n=1$ case of Equation \ref{equation:sersic_profile} and is parameterized using its (source-plane) center, effective radius, and intensity. With higher-resolution imaging data, the background source light profile could be fit with a more detailed model. The background source center coordinates are initialized to values within 2 arcseconds of the line of sight of the foreground lens center. The intensity of the background source light is again set to a wide log-uniform prior distribution, as for the lens light in Search 1. The source effective radius is initialized with a Gaussian distribution around a typical disk galaxy size, as discussed in Section \ref{sect:priors}. To map coordinates to the source plane, the lens galaxy's total mass is modeled as a singular isothermal elliptical (SIE) profile. The mass profile's Einstein radius prior is a wide Gaussian distribution centered at 1.0 arcsecond with a hard upper limit of 2.5 arcseconds.  The center and elliptical components of the SIE are paired with the light profile with the assumption that the ellipses will be aligned. The lens light profile takes prior distributions passed from the results of Search 1. By fixing the center of the lens profiles to the results of Search 1, the lens model in this search is reduced to 10 free parameters. This, in addition to taking informed Gaussian priors from Search 1 for the lens light, helps the model to focus on solutions that fit the source-light instead of systematic solutions that fit artefacts in the data. 
We use the annular \textit{source} mask that removes the lens light from the observed image and therefore further focuses the model on fitting the source light. We also utilize {\sc PyAutoLens}'s position resampling functionality, whereby the brightest pixels in the lensed source are marked (via a GUI). Again, the results of this search are passed as priors to Search 3.

\subsection{Search 3 --- Combined Lens and Source Models}

Search 3 fits every component of the system. To model the foreground lens, we use a combined elliptical S\'ersic mass-light profile for the stellar component and an elliptical NFW profile for the dark matter halo. Background source light is modeled again as a spherical exponential profile. Priors are passed from Search 2 (see Table \ref{table:phases}), except for the dark matter halo profile and stellar mass-to-light ratio. The prior distributions for these crucial parameters are determined by calculating central and limiting values according to a more careful process, as described in Section \ref{sect:priors}. The \textit{full} mask including all the lens and source features is used to remove background features and other contaminants that exist in the periphery, which saves computational time. Lensed image positions are again used to discard unphysical mass models. This final search produces a reasonable fit to the complexities introduced by each component and gives uncertainties on each of the inferred quantities. Additional disk and bulge components, cores, and multiple galaxies at different planes along the line of sight can be fitted and would allow more precise and realistic models. These improvements to realism are unhelpful here given the quality of imaging available for the objects in question but would be simply applied in future studies following the same principles outlined in our strategy.


\section{Highest-Quality Model Results}
\label{sect:other_models}

\begin{figure*} 
    \includegraphics[width=\columnwidth]{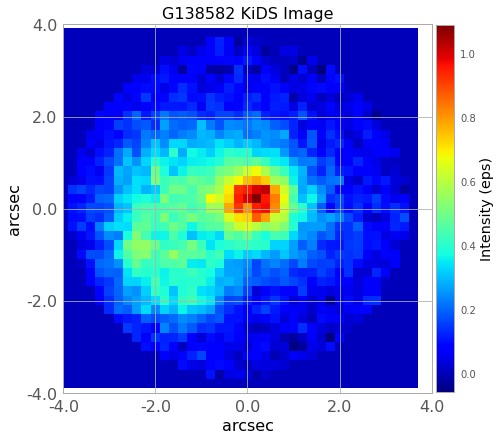}
    \includegraphics[width=\columnwidth]{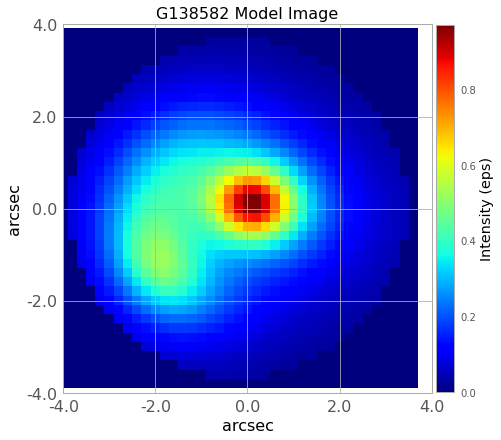}
    \includegraphics[width=\linewidth, trim={3.4cm 0 4.0cm 0}, clip]{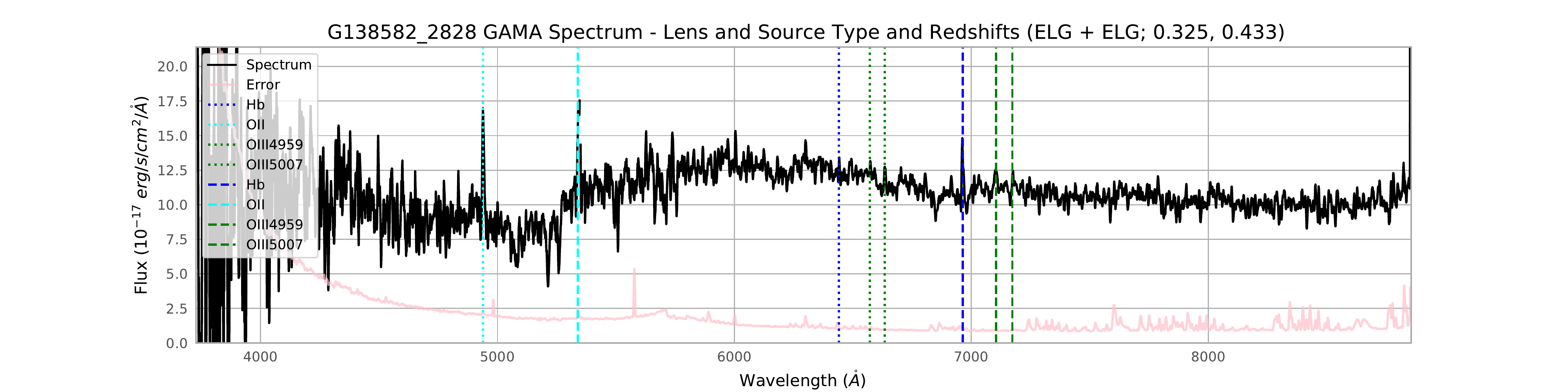}
    \includegraphics[width=\linewidth, trim={3.4cm 0 4.0cm 0}, clip]{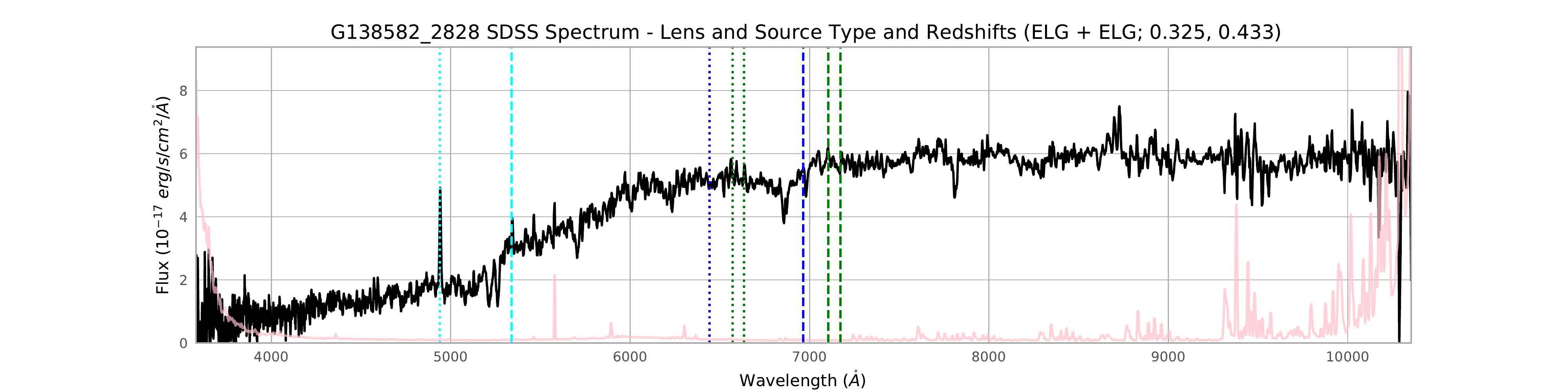}     
    \caption{G138582. A-Grade. \textit{Upper left}: The observed image shows a single bright elongated feature in the lower left of the foreground lens galaxy profile with a tail in the upper part of the feature. \textit{Upper right}: The model image correctly captures the shape of the image with lensing characteristics. \textit{Lower}: The GAMA and SDSS spectra both show reasonably strong emission lines (H$\beta$, [O II], [O III]) for the foreground lens galaxy at $z = $ 0.325 (dotted) and the background source galaxy at $z = $ 0.433 (dashed). Foreground lens CaH\&K absorption lines are also easily identified but are left ummarked here to show the presence of emission lines for this spectrum's ELG+ELG {\sc AUTOZ} template match.}
    \label{fig:2828}
\end{figure*}

\begin{figure*}
        \includegraphics[width=\columnwidth]{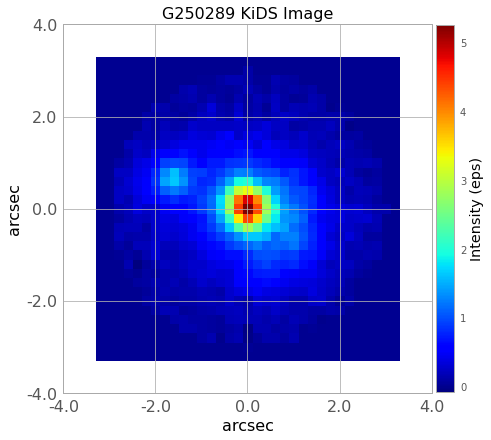}
        \includegraphics[width=\columnwidth]{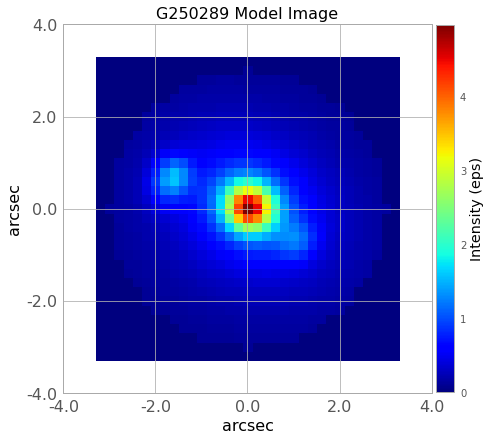}
        \includegraphics[width=\linewidth, trim={3.4cm 0 4.0cm 0}, clip]{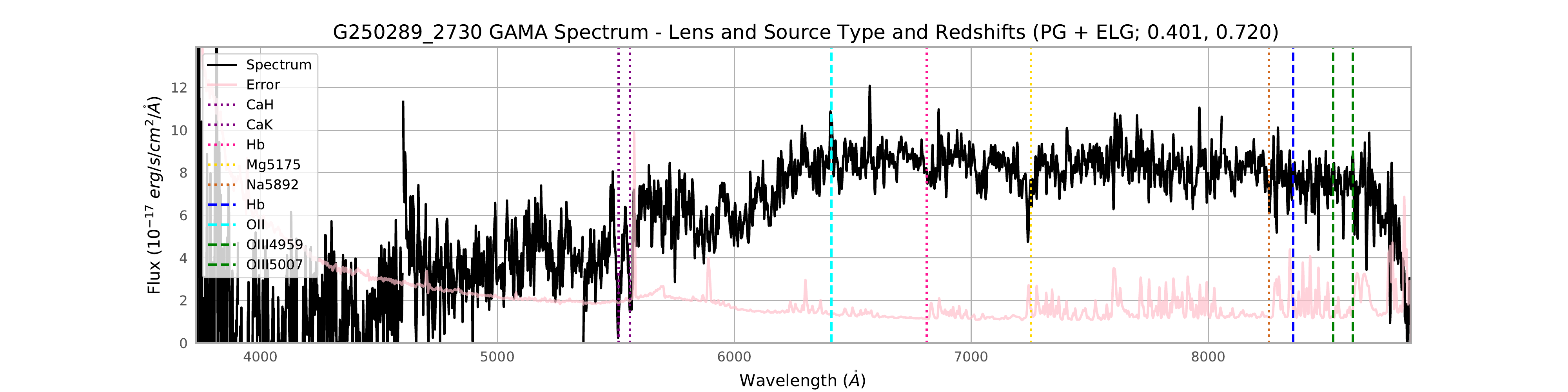}
        \includegraphics[width=\linewidth, trim={3.4cm 0 4.0cm 0}, clip]{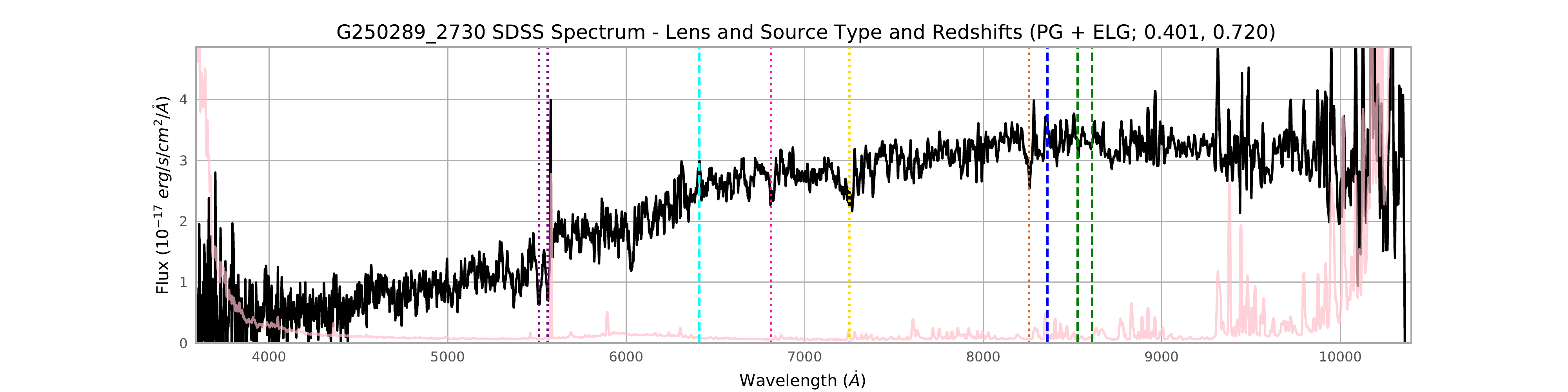}
        \caption{G250289. B-Grade. \textit{Upper left}: The observed image shows a doubly imaged source with a near-elliptical shape in the upper left with respect to the foreground lens and an arc mirrored across the lens to the lower right that blends somewhat with the foreground lens light. \textit{Upper right}: The model reconstructs the locations of both images, but the mirrored image in the lower right lacks the stretched elongated shape, which may be an effect of internal structure that is unaccounted for in the model. \textit{Lower}: The absorption features shown with dotted lines (CaH, CaK, H$\beta$, Mg, and Na) of the foreground lens galaxy at $z = $ 0.401 are particularly strong. Emission of [O II] from the source galaxy at $z = $ 0.720 appears in both spectra with dashed lines, as well as weaker features from H$\beta$ and [O III]. However, the SDSS spectrum appears to show a stronger [O III]$\lambda$4959 than [O III]$\lambda$5007, which should not be the case. The weak background source-flux could be because much of the the upper left source feature in the observed image is outside the 1- and 1.5-arcsecond GAMA and SDSS apertures. }\label{2730}
\end{figure*}

\begin{figure*}
    \includegraphics[width=\columnwidth]{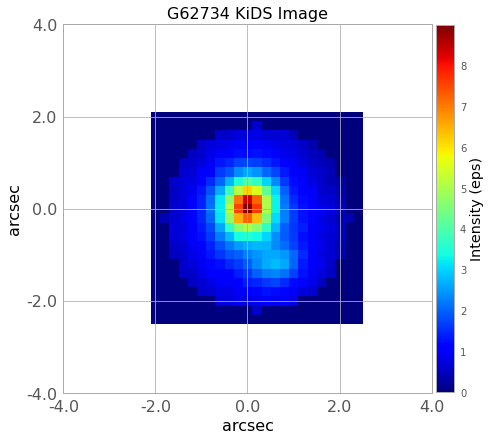}
    \includegraphics[width=\columnwidth]{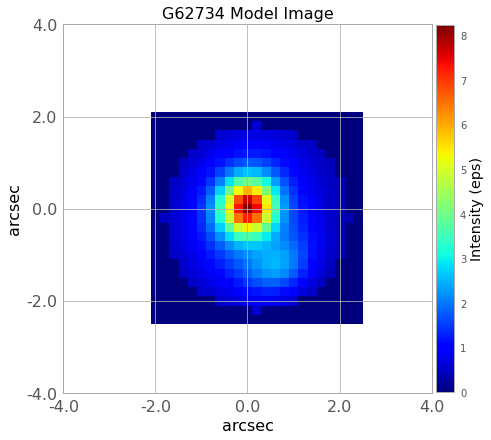}
    \includegraphics[width=\linewidth, trim={3.4cm 0 4.0cm 0}, clip]{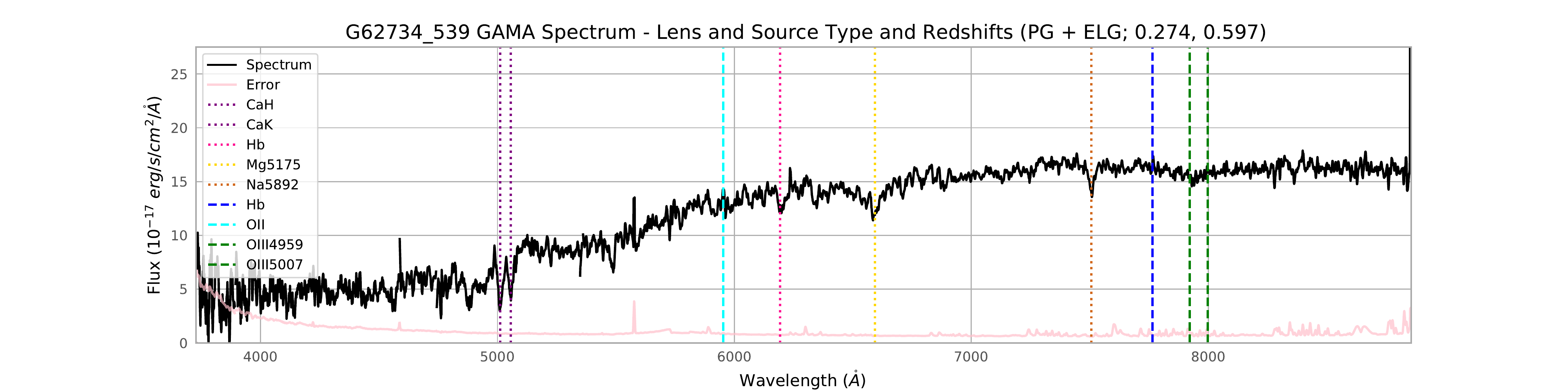}
    \caption{G62734. B-Grade. \textbf{Dark mass poorly constrained, so not included in further analysis alongside the other A- and B-grade models.} \textit{Upper left}: The observed image shows an image in the lower right with respect to the foreground lens profile. \textit{Upper right}: The shape and location of the lensed source feature in the lower right are well-fit in the model image, but there is some extra light surrounding the lensed source feature that may be due to the less-sophisticated spherical exponential light profile that we use to model the source light. The exact reconstruction of the source light profile is not the main goal of this exercise, though higher resolution imaging would make it worth further constraining with more flexible priors. \textit{Lower}: Foreground lens absorption features (CaH, CaK, H$\beta$, Mg, and Na) are clearly shown with dotted lines at $z = $ 0.274, and weak emission features can be identified with dashed lines at $z = $ 0.597.} \label{fig:539}
\end{figure*}

\begin{figure*}
    \includegraphics[width=\columnwidth]{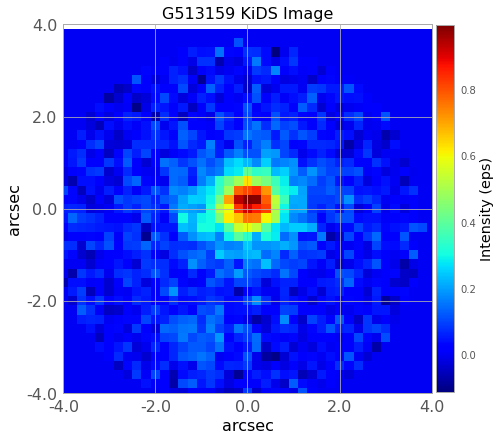}
    \includegraphics[width=\columnwidth]{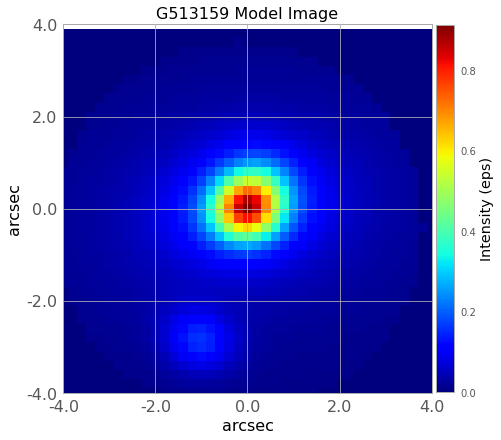}
    \includegraphics[width=\linewidth, trim={3.4cm 0 4.0cm 0}, clip]{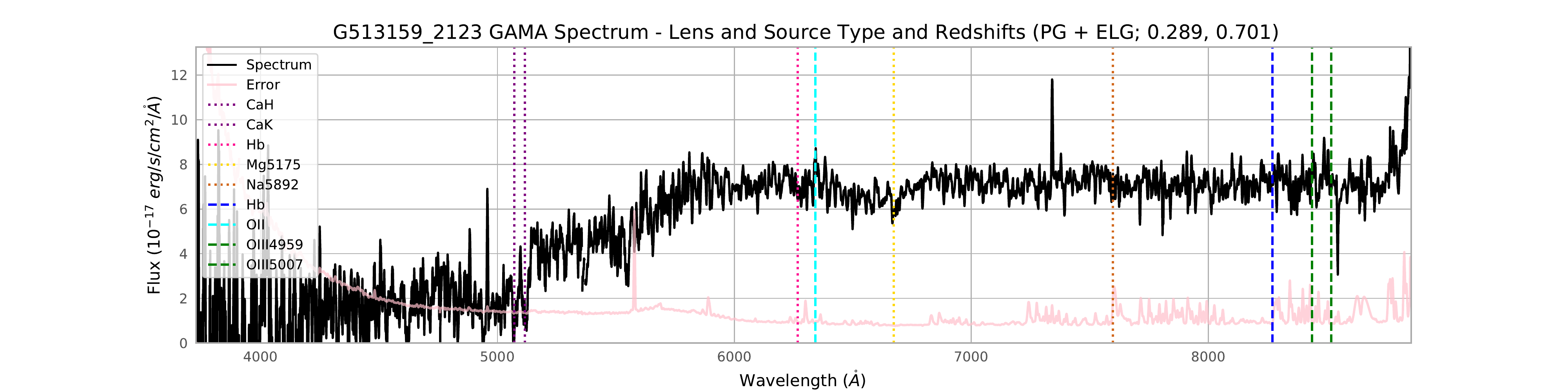}
    \caption{G513159. B-Grade. \textit{Upper left}: The observed image shows a feature around 3 arcseconds away from the central foreground lens profile that may be a lensed source feature. \textit{Upper right}: The model successfully accounts for the position and flux of the extra light through lensing. \textit{Lower}: The GAMA spectrum shows strong CaH and CaK features with dotted lines for the foreground lens galaxy at $z = $ 0.289 and possible emission line features ([O II] and [O III]) with dashed lines at $z = $ 0.701. Some expected features are plotted but not well-defined in the spectrum.}\label{fig:2123}
\end{figure*}

Figures \ref{fig:2828}-\ref{fig:2123} show the observed image, model image, and spectra for some of the most successful models.

\section{Spectrum Quality Control}\label{sect:spec_qual_ctrl}

We return to the specific cases described in Section \ref{sec:autoz_cases} to see how they affected our final quality scoring and subsample selection in the interest of retaining the most true positives while minimizing the inclusion of false positives.

\subsection{When there are Overlapping Emission or Absorption Lines...}\label{sect:sample_overlaps}

Almost half of the candidates (20 of 42) we selected initially by {\sc Autoz} output had overlapping line features of some kind in their spectrum. 12 of these were overlapping absorption features; 8 were emission features. 6 of the 19 candidates that were accepted following critical quality control had overlaps. The presence of an overlap affected the scoring of the individual spectrum, which was accepted only if the other background source line features were well-shown. We classify the cases of overlap as "on-template" or "off-template" in reference to the background source template. "Off-template" overlaps are cases when the overlapping line is an emission or absorption feature for a background source PG or ELG respectively, as opposed to "on-template" overlaps, where the overlap is emission or absorption for ELG or PG respectively. 9 of the 20 cases of overlap were "on-template". The other 11 were "off-template". The 6 overlapping cases that were retained in the final subsample of 19 graded models consisted of 1 on-template and 5 off-template overlaps.

7 of the 8 candidates with overlapping \textit{emission} features include background source ELGs (3 PG+ELG and 4 ELG+ELG), and one is a background source PG (ELG+PG). Two of these are retained in the 19 graded models. Recall from Section \ref{sec:autoz_cases} that all emission line overlaps are between the lens [O III]$\lambda\lambda$4959,5007 couplet and the source [O II]$\lambda3727$. It appears that these emission lines can have a significant effect even when one of the templates is a PG.
Of the 12 \textit{absorption} feature overlaps, 8 were PG+ELG, 2 were ELG+ELG, and 2 were PG+PG. 5 of these off-template PG+ELG absorption line overlaps make our final selection of 19 candidates, with a grade B, two C's, and two D's. One of the highest-scoring candidates (G250289, PG+ELG) had an overlap of foreground-lens H$\beta$ and background-source CaK absorption features, but the emission lines from the background source were well-defined and gave confidence to the redshift determination. This case is a bit odd considering the background galaxy was fit to an ELG template and would presumably be most heavily weighted by emission features. None of the ELG+ELG or PG+PG configurations with overlaps were accepted. One of the two candidates noted in Section \ref{sect:autoz-z2} that were removed from the  \cite{Holwerda15} sample had overlapping features. The other had a reasonable spectrum and failed for other reasons.

\subsection{When the Lens is described as an Emission Line Galaxy...} \label{sect:elg_lens}

\begin{figure}
    \centering
    \includegraphics[width=\columnwidth]{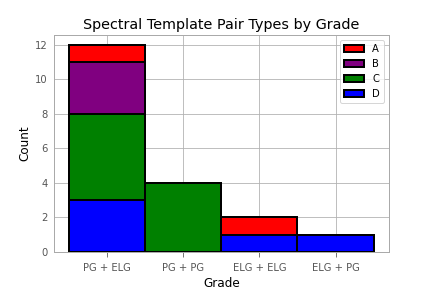}
    \caption{Stacked histogram of the four possible configurations of PG and ELG, written as foreground+background, separated by their quality grade (A, B, C, D) as described in Section \ref{sect:quality_scoring}. The large majority of successful models were composed of a passive foreground lens galaxy and emission line background source galaxy, which is expected. Other configurations are less likely, but one of the two A-grade models came from an ELG+ELG configuration.}
    \label{fig:grades_types}
\end{figure}

As discussed in Section \ref{sec:autoz_cases}, the case of an emission line galaxy acting as the foreground lens is less likely than a case where a passive galaxy acts as the lens. Only 3 ELG+ configurations were retained in the graded subsample of 19 candidates, two of which were low-scoring D-grades. Only one of the ELG+PG configurations was accepted and given a D-grade. Figure \ref{fig:grades_types} shows the foreground+background configurations for the 21 candidates in the final selection in the same manner shown in Figure \ref{fig:template_types}, now with quality grades. A, B, C, and D grades are blue, green, purple, and red respectively. Interestingly, one of the highest scoring (A-grade candidate G138582) candidates was one of the ELG+ELG matches. As shown in Appendix Figure \ref{fig:2828}, the emission lines from lens and source are clearly determined, and the resulting model was one of the most successful of this study. This example highlights the potential value of including (though with critical evaluation) the ELG+ foreground lens template configurations in the selection. Still, the template configurations shown in Figure \ref{fig:grades_types} mostly reaffirm the validity of the assumption that passive large elliptical galaxies provide the clearest and most usable foreground lenses. Further, since more background source +ELG template configurations have higher scores relative to +PG configurations, this again shows that the flux from strong emission lines in the background source is more detectable than the continuum and absorption features of a passive galaxy.

\subsection{When Source Emission Lines are Redshifted Beyond Observed Wavelength Range...}

27 of the initial 42 {\sc Autoz} spectra had +ELG configurations (i.e. background source is an emission line galaxy), 8 of which had H$\beta$ and [O III]$\lambda\lambda$4959,5007 emission lines redshifted beyond the GAMA upper wavelength limit of 8850{\AA}. These features would be present in the longer-wavelength upper range of the SDSS-BOSS spectrum for all 27 +ELG candidates, but not all were measured in SDSS-BOSS. 3 of the 19 candidates had all three above line features redshifted beyond the survey upper wavelength limit. Because these 3 objects were also measured with SDSS-BOSS spectroscopy and included in GAMA-DR3 {\sc SpecAll}, their emission lines redshifted beyond 8850{\AA} were detectable, but the AUTOZ match did not have access to those wavelengths. These were 2 C-grades and a D-grade.

\subsection{When Primary Redshift is Background Source...}\label{sect:primary_background}

10 of the initial 42 {\sc Autoz} spectra featured higher cross-correlation peaks to the background source than to the foreground lens (i.e. $\sigma_1$ is the match to the background source). 2 of those are included in the final graded 19 models. These two cases are shown in Table \ref{table:primary_background}. One of these is one of the two highest-scoring candidates (A-grade, candidate G323152, PG+ELG). G323152 represents the case described in the section \ref{sec:autoz_cases} where very strong emission lines from the background source are interpreted as the primary redshift match instead of the lower redshift passive continuum. The other candidate with $\sigma_1$ assigned to the background source flux is a D-grade with ELG+PG configuration. As mentioned before, we expect this configuration with the primary match to the background source redshift to be far less likely. Still, as with the ELG+ELG matches discussed in the previous section, the A-grade example of this case reinforces the value of including the {\sc Autoz} configurations where $\sigma_1$ is at higher redshift than $\sigma_2$.

\subsection{Additional Curiosities, Overlaps, Failures of our Utilization of AUTOZ}

Two of the ELG+PG configurations show the lens [O III]$\lambda$4959 line straddled by the H and K lines of the background source PG. This is a case where a "peak" between the two absorption valleys can be mistakenly considered an emission line feature. These are 2 of 4 foreground lens redshift matches below z$\sim$0.1. The other two have overlaps between lens [O III]$\lambda$5007 and source [O II]$\lambda3727$. A revision of the initial selection strategy could have extended the redshift cutoff to z$\sim$0.1 with no change to the sample. Two others appear to have emission lines fairly close to absorption lines, which might also give the impression of a peak or valley where it actually does not exist. One of these was accepted in the 19 and was given a grade of D.

\section{Subsample Selection and Context}
\label{s:kids:2zs}

We find that the majority of machine learning candidates did not pass our selection criteria covering {\sc Autoz} output parameters. This is predictable in light of the results of \cite{Knabel20}.

From the 421 LinKS candidates in the GAMA equatorial regions, there are 348 matching {\sc Autoz} entries (including duplicates) for 300 unique LinKS candidates. 59 of these entries pass the $R\geq1.2$ criterion, and 56 of these have galaxy-galaxy template matches. Four of those entries are duplicates, leaving 52 candidates \citep[including 6 from][]{Knabel20}. We remove 12 of these through our redshift criteria, which leaves 42 (40 unique) LinKS {\sc Autoz} foreground+background redshift matches. The two duplicates are shown in Table \ref{tab:duplicates}, with the accepted matches in bold text. For G544226, both entries show a PG template match at redshift $z=0.227$ with an ELG at redshift $z=0.650$. The accepted entry shows higher $\sigma_1$, $\sigma_2$, and R, and it attributes the primary redshift match to the foreground lens galaxy. The other entry is an example where $\sigma_1$ can refer to the background source galaxy and $\sigma_2$ to the foreground lens galaxy, effectively reversing which shows "better" match while still identifying the redshifts and type correctly. Note that $\sigma_1$ and $\sigma_2$ for the rejected entry are quite close (6.294 and 6.410 respectively). Both entries for the other duplicate candidate show the same primary match. The entry that is rejected has a secondary match to an ELG template at much closer redshift, which is most likely a false match. We remove the one LinKS candidate with a low redshift success probability and are left with 39 LinKS {\sc Autoz}-selected candidates, six of which were included in the final LinKS candidate selection of \cite{Knabel20}.

\begin{table}
    \centering
    \begin{tabular}[width=\textwidth]{c|c|c|c|c|c|c|}
         GAMA ID & Type & $z_1$ & $z_2$ & $\sigma_1$ & $\sigma_2$ & R \\ \hline
         \textbf{G544226} & \textbf{PG+ELG} & \textbf{0.227} & \textbf{0.650} & \textbf{9.393} &     \textbf{7.240} & \textbf{2.122} \\
         & PG+ELG & 0.650 & 0.227 & 6.294 & 6.410 & 0.650 \\
         \textbf{G262874} & \textbf{ELG+ELG} & \textbf{0.386} & \textbf{0.859} & \textbf{6.222} & \textbf{3.422} & \textbf{1.217} \\
          & ELG+PG & 0.386 & 0.195 & 9.339 & 4.817 & 1.416 \\\hline
    \end{tabular}
    \caption{Duplicate {\sc Autoz} entries for LinKS lens candidates. Boldface text indicates the selected entry. Type refers to foreground+background template matches. $z_1$ and $z_2$ refer to redshift matches corresponding to {\sc Autoz} cross-correlation peaks $\sigma_1$ and $\sigma_2$. R is a parameter that weights $\sigma_2$ to third and fourth matches.}
    \label{tab:duplicates}
\end{table}

32 of 48 Li-BG candidates in the GAMA equatorial fields have a match in {\sc Autoz}, with 53 entries including duplicates. 8 candidates (with no duplicates) are selected by the $R$ criterion. 5 of those 8 candidates are removed by our redshift criteria, leaving 3 unique candidates for analysis. 
One GalaxyZoo candidate has a match in the {\sc Autoz} catalog, but it does not pass selection criteria for followup. 

\begin{figure}
    \centering
    \includegraphics[width=\columnwidth]{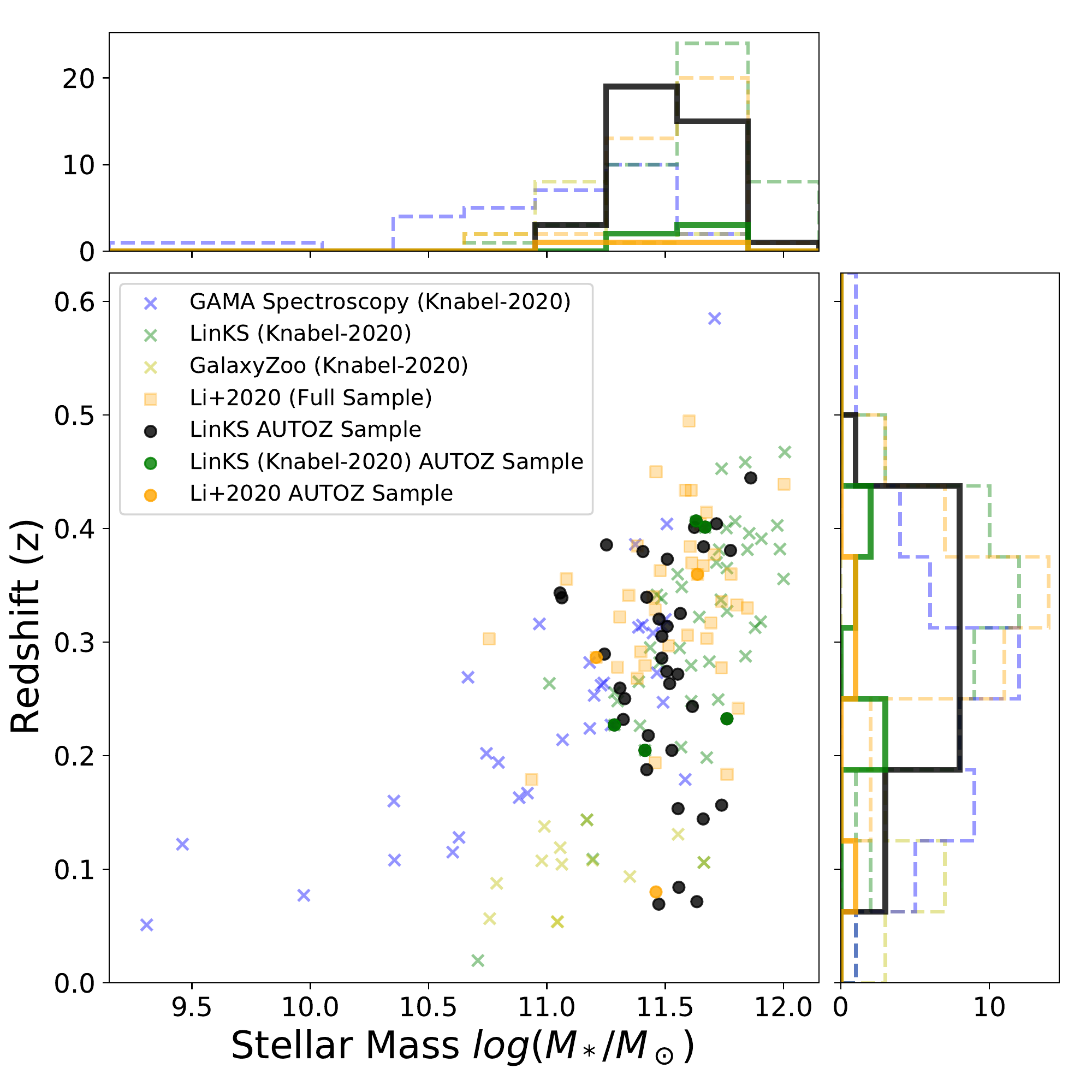}
    \caption{Stellar masses and redshifts of the {\sc Autoz} sample with deeper colored circular markers shown against the candidates discussed in \citealp{Knabel20} with faded X's for context. LinKS candidates (shown in green for the LinKS subsample selected in \citealp{Knabel20} and black for those that were not) and "bright galaxy" candidates from \citealp{Li20} (orange) have high stellar masses at intermediate redshift $log(M_*/M_{\odot})\sim$11-11.75 at $z\sim$0.2-0.5. Blue and yellow X's are spectroscopy and citizen-science candidates selected in \citealp{Knabel20}.}
    \label{fig:z_mass_comparison}
\end{figure}

In order to briefly contextualize this selection in reference to some of the results and conclusions drawn in \cite{Knabel20}, we show the {\sc Autoz} sample of 42 candidates selected in this work in Figure \ref{fig:z_mass_comparison} with circular markers in comparison with the candidates discussed in \cite{Knabel20} shown in the background with X's. Stellar mass estimates and lens redshifts shown here are from GAMA-DR3 {\sc StellarMassesLambdar} catalog. The {\sc Autoz} sample is slightly lower in stellar mass on average than the LinKS subsample as selected in \cite{Knabel20}, with a mean and median $\log{M_*}$ of (11.50, 11.51) compared to (11.61, 11.67). A Kolmogorov-Smirnov test of the stellar masses between the LinKS {\sc Autoz} subsample and the LinKS subsample as selected in \cite{Knabel20} results in a KS-metric of 0.352 with a p-value of 0.007, indicating a statistically significant disparity between the masses of the two selections. In fact, when compared to the GAMA spectroscopy subsample as selected in \cite{Knabel20}, the KS-test results are almost identical (metric 0.353, p-value 0.007). The bulk of {\sc Autoz} candidates hovers in the parameter space overlapping the upper mass end of the GAMA spectroscopic candidates and the lower mass end of the LinKS from \cite{Knabel20} candidates, which is reasonable if they are to be large enough to have distinguishable features for identification by machine learning while being small enough to have a higher chance of flux from the lensing features being collected in the 1-arcsecond GAMA spectroscopic fiber aperture. 

Two candidates in the {\sc Autoz} sample have $\sigma_2$ and \textit{R} values that would place them in the selection space defined for the \cite{Holwerda15} blended spectra candidates. One of them (G184530) was not selected in that study because it is an ELG+PG configuration (i.e. the emission line match is at closer redshift). The other (G544226) was removed because \cite{Holwerda15} removed candidates near the alias of $(1+z_1)/(1+z_2)=1.343\pm0.002$, corresponding to an overlap between redshifted [O II]$\lambda3727$ and [O III]$\lambda5007$ emission lines. G544226 then would have been the one overlap between the GAMA spectroscopic and LinKS machine learning catalogs in \cite{Knabel20} if it had not been removed. G544226 made the selection for high-quality candidates in \cite{Knabel20}. With a redshift of $z=0.227$ and $\log{M_*}=11.29$, it existed squarely in the overlap of parameters space between the GAMA spectroscopy and LinKS machine learning candidates. 

\bsp	
\label{lastpage}
\end{document}